\documentclass{JHEP3}
\usepackage{latexsym}

\makeatletter

\newcommand{\nn}{\nonumber\\}\newcommand{\p}[1]{(\ref{#1})}

\newcommand{\be}{\begin{equation}}
\newcommand{\ee}{\end{equation}}
\newcommand{\bea}{\begin{eqnarray}}
\newcommand{\eea}{\end{eqnarray}}
\newcommand{\lb}{\label}
\def\a{\alpha}\def\e{\epsilon}

\def\om{\omega}
\makeatletter
 \@addtoreset{equation}{section}

 \makeatother

\title{
\begin{flushright}
{{\small
{DFPD 02/TH/07, IFIC-02-16, FTUV/02-1904\\
 hep-th/0205104
}}}
\end{flushright}
~\\
{\bf On the Superconformal Flatness of AdS Superspaces}
}
\bigskip
\author{
 Igor Bandos~$^{1,2}$\thanks{bandos@ific.uv.es},
 Evgeny Ivanov${~}^3$\thanks{eivanov@thsun1.jinr.ru},
 Jerzy Lukierski${~}^4$\thanks{lukier@ift.uni.wroc.pl}
 ~and
 Dmitri Sorokin${~}^{1,5}$\thanks{sorokin@pd.infn.it}\\
~\\
$^1${\it\small Institute for Theoretical Physics, NSC KIPT, 61108 Kharkov, Ukraine}\\
 $^2${\it\small Departamento de F\'{\i}sica Te\'orica and IFIC
(CSIC-UVEG), 46100-Burjassot (Valencia), Spain} \\
$^3${\it\small  Bogoliubov Laboratory of Theoretical Physics, JINR,}\\
{\it \small 141980 Dubna, Moscow Region, Russian Federation}\\
$^4${\it\small  Institute for Theoretical Physics, University of Wroclaw,
 50-204 Wroclaw, Poland}\\
$^5${\it\small Universit\`a degli Studi di Padova,
Dipartimento di Fisica ``Galileo Galilei'',}\\
{\it\small INFN, Sezione di Padova, via F. Marzolo, 8, 35131 Padova,
Italia}}
\date{}

\maketitle
\newpage
 \abstract{ The superconformal structure of coset
superspaces with $AdS_m\times S^n$ geometry of bosonic subspaces is studied. It
is shown, in particular, that the conventional superspace extensions of the
coset manifolds $AdS_2\times S^2$, $AdS_3\times S^3$ and $AdS_5\times S^5$,
which arise as solutions of corresponding $D=4,6, 10$ supergravities and have
been extensively studied in connection with $AdS/CFT$ correspondence, are not
superconformally flat, though their bosonic submanifolds are conformally flat.
We give a group--theoretical reasoning for this fact. We find that in the
$AdS_2\times S^2$ and $AdS_3\times S^3$ cases there exist different supercosets
based on the supergroup $OSp(4^*|2)$ which are superconformally flat. We also
argue that in $D=2,3,4$ and 5 there exist superconformally flat `pure' $AdS_D$
supercosets. Two methods of checking the superconformal flatness are proposed.
One of them consists in solving the Maurer--Cartan structure equations and the
other is based on embedding the isometry supergroup of the $AdS_m\times S^n$
superspace into a superconformal group in $(m+n)$--dimensional Minkowski space.
Finally, we discuss some applications of the above results to the description
of supersymmetric dynamical systems.}

\begin{document}

\renewcommand{\thefootnote}{\arabic{footnote}}
\newpage
\setcounter{page}1

\section{Introduction}

Space--times of anti--de--Sitter geometry have attracted great deal of
attention because they appear in various physical problems, e.g. cosmology,
black holes, supergravity and compactification, $AdS/CFT$ correspondence, the
theory of higher spins, etc. The basic geometrical feature of the
$D$--dimensional $AdS$ spaces is that their isometry group $SO(2,D-1)$ acts as a
group of conformal transformations of the $AdS$ boundary which may be
identified with a $D-1$ dimensional Minkowski space. On the other hand, the
$AdS$ metric is invariant, up to a dilaton factor, under a higher group of
conformal transformations, namely $SO(2,D)$. A consequence of this fact is that
(locally) there exists a set of $AdS$ coordinates $x^m$ in which the $AdS$
metric is conformally flat\footnote{One can mention another peculiar property
of the $AdS$ metric to have (in a certain set of coordinates) a `Kahler--like'
potential structure $g_{mn}=\partial_m\partial_nV(x)$ \cite{kahler}.}
\begin{equation}\label{D1}
ds^2=e^{4\phi(x)}dx^mdx^n\eta_{mn}\,, \quad \eta_{mn}={\rm
diag}{(+1,-1\cdots,-1)}, \quad m,n=0,1,...,D-1\,,
\end{equation}
where $\phi(x)$ is a conformal factor.

Among compactifications of $D=10$ and $D=11$ supergravities (for reviews see
\cite{kk}) there appear vacuum configurations having the geometry of the direct
product of an $AdS$ space and a sphere, i.e. $AdS_p\times S^q$ ($p+q=D$). These
configurations provide a geometrical ground for the $AdS/CFT$ correspondence in
string theory and M--theory \cite{ads/cft}, and also are relevant (for $p=q=2$)
to the superconformal quantum mechanics \cite{qm} of particles in the
background of Reissner--Nordstr\"om black holes \cite{black}. The isometry
group of such spaces is $SO(2,p-1)\times SO(q+1)$ which is a bosonic subgroup
of the appropriate superconformal symmetry of the given configuration of string
or M--theory on the boundary of $AdS$.

It is known that the $AdS_5\times S^5$ space of the compactified type IIB
$D=10$ supergravity (or string theory) and the $AdS_2\times S^2$ space of
Reissner--Nordstr\"om black hole (see e.g. \cite{black,cflat}) are conformally
flat since the radii of the $AdS$ spaces and spheres are adjusted
to be equal. This is also the case for superstrings and superparticles in
$AdS_2\times S^2$ and $AdS_3\times S^3$ \cite{pesando,berkovits,zhou}, while
the compactifications of $D=11$ supergravity on $AdS_4\times S^7$ and
$AdS_7\times S^4$ are not conformally flat. Conformal flatness of classical
solutions has been considered as a guarantee of their exactness, i.e. the
absence of quantum corrections to these solutions due to higher derivative
terms in the supergravity effective actions \cite{bg}. As has been discussed in
\cite{kallosh1}, since the $AdS_4\times S^7$ and $AdS_7\times S^4$ solutions of
$D=11$ supergravity are not conformally flat, the exactness argument for them
should be based on another reasoning, such as unbroken supersymmetry. In this
connection one can also ask whether the conformal flatness is compatible with
the supersymmetry properties of the corresponding configurations.

 So a natural question arises whether the
 {\it superspaces} with a conformally
 flat bosonic body $AdS_p\times S^q$ are also {\it superconformally}
 flat. A naive expectation might be that
 the answer to this question is always positive. However, we shall see
 that, among physically interesting examples, indeed the $AdS_4$ coset
superspace ${{OSp(N|4;R)}\over{SO(1,3)\times SO(N)}}$ (for a generic $N)$ is
superconformally flat, while the $AdS_2\times S^2$ coset superspace
${{SU(1,1|2)}\over{SO(1,1)\times SO(2)}}\,,$ the $AdS_3\times S^3$ coset
superspace ${{SU(1,1|2)\times SU(1,1|2)}\over{SO(1,2)\times SO(3)}}$ and the
$AdS_5\times S^5$ coset superspace ${{SU(2,2|4)}\over{SO(1,4)\times SO(5)}}$
are not, though their bosonic subspaces are conformally flat. The reason for
this somewhat surprising conclusion is that the isometry supergroups
$OSp(N|4;R)$ of the corresponding four--dimensional $AdS$ superspaces are {\it
subgroups} of the $N$--extended $D=4$ superconformal group
$SU(2,2|N)\,,$\footnote{The Cartan forms of the N extended superconformal group
$SU(2,2|N)$ which are suitable for the construction of $AdS_5$ coset
superspaces, were computed in \cite{abz}. We also note that the bosonic
subgroup of $SU(n,n|2n)$ $(n=1,2)$ is $SU(n,n)\times SU(2n)\,,$ while for a
generic $SU(n,n|N)$ it is $SU(n,n)\times U(N)\,.$} like the super Poincar\'e
group of flat $N$--extended $D=4$ superspaces. On the contrary, the isometry
supergroups $SU(1,1|2)$, $SU(1,1|2)\times SU(1,1|2)$ and $SU(2,2|4)$ of the
$AdS_{D\over 2} \times S^{D\over 2}$ (D=4,6,10) superspaces are not {\it
appropriate} subgroups of corresponding $N=2$, $D=4,6,10$ superconformal
groups. By `appropriate' we mean that the bosonic subgroups of these $AdS$
supergroups must be subgroups of the bosonic conformal subgroups of the
corresponding  superconformal groups. This is the group--theoretical argument
why the above mentioned coset superspaces are not superconformal, though their
bosonic $AdS_{D\over 2} \times S^{D\over 2}$ subspaces are
conformal.\footnote{Let us note that (super)conformally flat (super)spaces are
(super)conformal in the sense that, with a proper choice of
coordinates, the whole (super)conformal group is realized
in these (super)spaces in the same way as in the corresponding
flat (super)spaces. So below we shall often use `(super)conformal'
as a synonym of  `(super)conformally flat'.}

For instance, the isometry supergroup $SU(2,2|4)$ of the $AdS_5 \times S^5$
solution of type IIB $D=10$ supergravity and the type IIB $D=10$ super
Poincar\'e group are {\it not subgroups} of a generalized  extended (`$N=2$')
superconformal group in ten dimensions (usually chosen to be either
$OSp(2|32;R)$ or $OSp(1|64;R)$). At the same time, the bosonic $D=10$
Poincar\'e group and the isometry group $SO(2,4)\times SO(6)$ of $AdS_5\times
S^5$ are subgroups of the $D=10$ conformal group $SO(2,10)$.

As was already discussed in the literature \cite{osp164,bars,west}, only
`central' extensions of the type IIB $D=10$ super Poincar\'e group by tensorial
generators are subgroups of $OSp(2|32;R)$ and/or $OSp(1|64;R)$,\footnote{These
central extensions nontrivially transform under the superconformal
transformations.} while $SU(2,2|4)$  is {\it not} a subgroup of these
superconformal groups (we give a simple reasoning for this in Subsection 2.4).
 As a result, the $D=10$ super Poincar\'e group and the $SU(2,2|4)$ supergroup
and, respectively, the flat $D=10$ superspace and the $AdS_5\times S^5$
superspace, cannot be related to each other by a super Weyl transformation (in
the sense explained in Sections 3 and 4). Hence the $AdS_5\times S^5$
superspace is not superconformal.

In the cases of $AdS_2\times S^2$ and $AdS_3\times S^3$, however, there exist
different coset superspaces with the same bosonic body which are
superconformal. These are ${{OSp(4^*|2)}\over{SO(1,1)\times SO(2)\times
SU(2)}}$ and ${{OSp(4^*|2)\times OSp(4^*|2)}\over{SO(1,2)\times SO(3)\times
SU(2)\times SU(2)}}\,.$ These superspaces contain $AdS_2\times S^2$ and
$AdS_3\times S^3$ as bosonic subspaces and have the same number of fermionic
coordinates as the supercosets ${{SU(1,1|2)}\over{SO(1,1)\times SO(2)}}$ and
${{SU(1,1|2)\times SU(1,1|2)}\over{SO(1,2)\times SO(3)}}\,,$ respectively.
Their crucial difference from the latter ones
is that the supergroups $OSp(4^*|2)$
and $OSp(4^*|2)\times OSp(4^*|2)$ are subgroups of the corresponding
superconformal groups $SU(2,2|2)$ and $OSp(8^*|4)\,.$ These coset superspaces are
related to the `conventional' ones by an `$\alpha$-deformation' of  supercosets
based on the exceptional supergroup $D(2,1;\alpha)$ (see Sections 2 and 4).

 A generic form of the supervielbeins of supermanifolds with
superconformally flat geometry is \cite{Howe}
\begin{eqnarray}\label{superflat}
E^a&=&e^{2\Phi(x,\theta)}(dx^a-i\bar\theta\gamma^ad\theta)\equiv
e^{2\Phi(x,\theta)}\Pi^a,
\nonumber\\
E^{\alpha}&=&e^{\Phi(x,\theta)}\Lambda^{\alpha}_{~\beta}(d\theta^\beta
+ iD_\gamma\Phi \gamma_a^{\gamma\beta}\Pi^a)\,,
\end{eqnarray}
where $x^m$ and $\theta^\alpha$ are bosonic and
fermionic coordinates of the supermanifold, $\Phi(x,\theta)$ is a
dilaton factor, $\Lambda^{\alpha}_{~\beta}(x,\theta)$ is a matrix
(in some cases it can be equal to unity), $D_\alpha$ is a flat
superspace covariant derivative, and $\Pi^a$ and $d\theta^\alpha$
are covariant flat supervielbeins.

In this paper we shall derive the superconformally flat form of the
supervielbeins and connections of the $AdS_4$ coset superspaces
${{OSp(N|4;R)}\over{SO(1,3)\times SO(N)}}$ $(N=1,2)$ and of an $AdS_2\times
S^2$ coset superspace ${{OSp(4^*|2)}\over{SO(1,1)\times SO(2)\times SU(2)}}\,.$
Actually, the first type of superspaces will be shown to be superconformally
flat for generic $N$. In Section 2 we show that the superconformally flat
ansatz is compatible with the Maurer--Cartan structure equations of the
corresponding coset superspaces (while it is not compatible with the $AdS\times
S$ solutions of $N=2$, $D=4,6$ supergravities), and in Sections 3 and 4, using
a `bottom--up' approach, we construct explicitly the superconformal factors of
the $AdS$ supervielbeins. {In Subsection 2.4 we demonstrate that the supercoset
extension of $AdS_5\times S^5$ is not superconformally flat}. In Conclusion we
also discuss the issue of the supeconformal flatness of 'pure' $AdS_D$
supercosets, i.e. those without `$S$--factors'. In particular, we argue that
there exists only one superconformlly flat `pure' $AdS_5$ supercoset, with
$SU(2,2|1)$ as the superisometry group.

The results obtained can be useful for the description of supersymmetric black
holes, superparticles and superstrings in $AdS$ superbackgrounds, and for
studying issues of their quantization and the influence of higher order quantum
corrections. As an example, in Section 5 we demonstrate that the classical
dynamics of massless superparticles propagating on the $AdS_4$ coset superspace
${{OSp(2|4;R)}\over{SO(1,3)\times SO(2)}}\,,$ on the $AdS_2\times S^2$ coset
superspace ${{OSp(4^*|2)}\over{SO(1,1)\times SO(2)\times SU(2)}}$ and in flat
$N=2$, $D=4$ superspace are equivalent, because these superspaces are
superconformal, with the same superconformal group $SU(2,2|2)$ acting in all
three cases. But their quantized dynamics differ because of different
geometrical and symmetry properties of the superbackgrounds which the
quantization procedure should respect.

\section{Solving the Maurer--Cartan equations of the AdS
cosets}

In this section we use two-component Weyl spinors $\theta^\alpha$,
$\bar\theta^{\dot\alpha}$ to define Grassmann coordinates and the matrix
representation $x^{\alpha\dot\alpha}=\sigma_a^{\alpha\dot\alpha}x^a$ for the
vector coordinates of $D=4$ superspaces. The metric of the flat space--time is
chosen to be almost negative, $\eta_{ab}=\mbox{diag}(+,-,-,-)$.

\subsection{AdS$_4$ superspaces}

The structure equations of the $AdS_4$ coset superspace
${OSp(1|4;R)}\over{SO(1,3)}$ are (see e.g. \cite{isor}, \cite{bls})
 \begin{equation}\label{DEa}
    {\cal D}E^{\alpha\dot{\alpha}} =
    dE^{\alpha\dot{\alpha}}
    -E^{\beta\dot{\alpha}} \wedge w_\beta^{~\alpha}
    -E^{\alpha\dot{\beta}} \wedge w_{\dot{\beta}}^{~\dot{\alpha}} =
    -2i E^\alpha \wedge \bar{E}^{\dot{\alpha}},
\end{equation}
 \begin{equation}\label{DEA}
    {\cal D}E^{\alpha} =
    dE^{\alpha}
    -E^{\beta} \wedge w_\beta^{~\alpha}=
    - {1 \over R} \bar{E}_{\dot{\beta}}
    \wedge
    {E}^{\alpha\dot{\beta}},
\end{equation}
 \begin{equation}\label{R}
    R^{\alpha\beta} =
    dw^{\alpha\beta}
    -w^{\alpha\gamma} \wedge w_\gamma^{~\beta}=
    - {2i \over R} E^\alpha
    \wedge E^\beta +
     {1\over R^2}
    {E}^{\alpha\dot{\gamma}}
    \wedge {E}^\beta_{~\dot{\gamma}}\,,
\end{equation}
where $w^{\alpha\beta}$ and $R^{\alpha\beta}$ are the spin connection and the
curvature form, and $R$ is the $AdS$ radius. The equations
(\ref{DEa})--(\ref{R}) should be supplemented with their complex conjugate.

Let us assume that  the $N=1$ $AdS_4$ superspace is
superconformally flat. In this case eqs. \p{DEa}--\p{R} should
have a nontrivial solution of the form
 \begin{equation}\label{cf}
    {E}^{\alpha\dot{\alpha}}=
    e^{2\Phi (x,\theta, \bar{\theta})}
    \Pi^{\alpha\dot{\alpha}}~,
\end{equation}
where
 \begin{equation}\label{Pi}
    \Pi^{\alpha\dot{\alpha}}=
    dx^{\alpha\dot{\alpha}}
    - i d\theta^\alpha \bar{\theta}^{\dot{\alpha}}
    + i \theta^\alpha d\bar{\theta}^{\dot{\alpha}}~.
\end{equation}

Substituting \p{cf} into the structure equation
    \p{DEa}, one finds that it fixes the form of the fermionic supervielbeins
    $E^\alpha, \bar{E}^{\dot{\alpha}}$ up to
    a phase factor $e^{i W (x,\theta, \bar{\theta})}\,,$ where $W$ is
a real superfield:
 \begin{equation}\label{cfEA}
    {E}^{{\alpha}}=
    e^{\Phi (x,\theta, \bar{\theta})
   +i W (x,\theta, \bar{\theta})}
    \left(
    d{\theta}^\alpha + {2i}
    \Pi^{\alpha\dot{\alpha}}\bar{D}_{\dot{\alpha}}
     \Phi \right),
\end{equation}
 \begin{equation}\label{cfEdA}
    {E}^{\dot{\alpha}}=
    e^{\Phi (x,\theta, \bar{\theta})
   -i W (x,\theta, \bar{\theta})}
    \left(
    d\bar{\theta}^{\dot{\alpha}}- {2i}
    \Pi^{\alpha\dot{\alpha}}D_\alpha \Phi \right).
\end{equation}
 From eq. \p{DEa} one also finds the Grassmann components of the
connection 1-form
 \begin{equation}\label{cfw}
    {w}_\beta^{~\alpha}=
    -2 d\theta^\alpha D_\beta \Phi
    -2 d\theta_\beta D^\alpha \Phi +
    \Pi^{\gamma\dot{\gamma}}
    w_{\gamma\dot{\gamma}\beta}^{~~~~\alpha}~, \qquad w_{\dot{\beta}}^{~\dot{\alpha}}=
    (w_\beta^{~\alpha})^*~.
\end{equation}
 In eqs. \p{cfEA}--\p{cfw}
 \begin{equation}\label{DA}
    D_\alpha = \partial_\alpha + i \bar{\theta}^{\dot{\alpha}}
    \partial_{\alpha\dot{\alpha}}~, \qquad
    \bar{D}_{\dot{\alpha}} =
-\bar{\partial}_{\dot{\alpha}} -i{\theta}^{{\alpha}}
\partial_{\alpha\dot{\alpha}} \qquad
 \end{equation}
 are the flat superspace covariant spinor derivatives.

In order to obtain eqs. (\ref{cfEA})--(\ref{cfw}) one takes the covariant
differential of the
 bosonic supervielbein form \p{cf}
$$
    {\cal D} {E}^{\alpha\dot{\alpha}}= -2i e^{2\Phi}
    d\theta^{{\alpha}}\wedge
    d\bar{\theta}^{\dot{\alpha}}+ 2
    e^{2\Phi}
    \Pi^{\alpha\dot{\alpha}} \wedge d\Phi -
    e^{2\Phi}
    \Pi^{\alpha\dot{\beta}} \wedge
    \bar{\om}_{\dot{\beta}}^{~\dot{\alpha}} -
    e^{2\Phi}\Pi^{\beta\dot{\alpha}} \wedge \om_{{\beta}}^{~{\alpha}}
 $$
 and compares the result with the right hand side of \p{DEa}.

Let us substitute into the right hand side  of eq. \p{DEa} the most general
expression for the fermionic supervielbeins in terms of the independent
covariant 1-superforms
    $d\theta$, $d\bar\theta$ and $\Pi$
    $$
    E^\alpha =d\theta^\beta {\cal E}_\beta^{~\alpha} +
    d\bar{\theta}^{\dot{\beta}}
    \bar{{\cal E}}_{\dot{\beta}}^{~{\alpha}}
    + \Pi^{\beta\dot{\beta}} \tilde{\psi}_{\beta\dot{\beta}}^{~\alpha}~,
    $$
    $$
    \bar{E}^{\dot{\alpha}} =
    d\bar{\theta}^{\dot{\beta}}
    \bar{{\cal E}}_{\dot{\beta}}^{~\dot{\alpha}} +
    d\theta^\beta {\cal E}_\beta^{~\dot{\alpha}}
    + \Pi^{\beta\dot{\beta}}
    \tilde{\bar{\psi}}_{\beta\dot{\beta}}^{~\dot{\alpha}}~.
    $$
    Then, equating  the components of the terms containing the
    basic form $ d{\theta}^{{\beta}} \wedge d\bar{\theta}^{\dot{\beta}}$ on the
    left and right hand sides of eq. \p{DEa}, we have
    $$
    {\cal E}_\beta^{~\alpha}
    \bar{{\cal E}}_{\dot{\beta}}^{~\dot{\alpha}} +
    \bar{{\cal E}}_{\dot{\beta}}^{~{\alpha}}
    {\cal E}_\beta^{~\dot{\alpha}} =
    {\delta }_\beta^{~\alpha}
    {{\delta}}_{\dot{\beta}}^{~\dot{\alpha}} e^{2\Phi}~,
    $$
    while the components of the
    basic form $ d{\theta}^{{\beta}} \wedge d{\theta}^{{\gamma}}$
   give rise to the relations
    $$
    2{\cal E}_{(\beta}^{~\alpha}
    {\cal E}_{\gamma)}^{~\dot{\alpha}} \equiv
    {\cal E}_{\beta}^{~\alpha}
    {\cal E}_\gamma^{~\dot{\alpha}} +
    {\cal E}_{\gamma}^{~\alpha}
    {\cal E}_\beta^{~\dot{\alpha}} =0~.
    $$
    Under the assumption that
    ${\cal E}_{\beta}^{~\alpha}$ is invertible, the solution
    of the second equation is
    $$
    {\cal E}_\beta^{~\dot{\alpha}} =({\cal E}_\beta^{~\dot{\alpha}})^*=0~,
    $$
    then, from the first equation one gets
    $$
    {\cal E}_\beta^{~\alpha}
    \bar{{\cal E}}_{\dot{\beta}}^{~\dot{\alpha}}
    = {\delta }_\beta^{~\alpha}
    {{\delta}}_{\dot{\beta}}^{~\dot{\alpha}} e^{2\Phi} \quad \Rightarrow \quad
    {\cal E}_\beta^{~\alpha}
    = {\delta }_\beta^{~\alpha}
    e^{\Phi + i W}~,
    \quad
    \bar{{\cal E}}_{\dot{\beta}}^{~\dot{\alpha}}
    =
    {{\delta}}_{\dot{\beta}}^{~\dot{\alpha}} e^{\Phi - iW}~.
    $$

    Taking the above relations into account, one finds that the component of the basic form
    $\Pi^{\beta\dot{\beta}} \wedge d\bar{\theta}^{\dot{\gamma}} $
    in eq. \p{DEa} can be written in the form
 \begin{equation}\label{om1}
    \e_{\dot{\alpha}\dot{\beta}}
     \om_{\dot{\gamma} \beta\alpha} +
     \e_{\alpha\beta}
     \bar{\om}_{\dot{\gamma} \dot{\beta}\dot{\alpha}} =
     -2 \e_{\alpha\beta}\e_{\dot{\alpha}\dot{\beta}} \bar{D}_{\dot{\gamma}}
\Phi - 2i
     \e_{\dot{\alpha}\dot{\gamma}}
     \psi_{\beta\dot{\beta}~\alpha}\,,
\end{equation}
     where
     $
     \tilde{\psi}_{\beta\dot{\beta}~\alpha}= e^{\Phi + i W}
     \psi_{\beta\dot{\beta}~\alpha}
     $
     and the indices are `lowered' by  the unit antisymmetric matrix
     $\e_{\alpha\beta}\,,$
     $(\epsilon_{12}=-\epsilon_{21}=1)$, e.g.
     $
     V_{\alpha\dot{\alpha}}= \e_{\alpha\alpha^\prime }
     \e_{\dot{\alpha}\dot{\alpha}^\prime}
     V^{\alpha^\prime \dot{\alpha}^\prime}\,.$
     Now we should decompose the above equation  into the irreducible
     parts, using the fact that the spin connections are
     symmetric
     $$
     \bar{\om}_{\dot{\gamma} \dot{\beta}\dot{\alpha}} =
     \bar{\om}_{\dot{\gamma} \dot{\alpha}\dot{\beta}}~, \qquad
     \om_{\dot{\gamma} \beta\alpha} =
     \om_{\dot{\gamma} \alpha\beta}\,,
     $$
and decomposing the spin--tensor $ \psi_{\beta\dot{\beta}~\alpha}$
as follows
 $$
    \psi_{\beta\dot{\beta}~\alpha} =
    \bar{\psi}_{0\dot{\beta}} \e_{\alpha\beta}+
    \chi_{\dot{\beta}~\beta\alpha}~, \qquad
    \chi_{\dot{\beta}~\beta\alpha} =
    \chi_{\dot{\beta}~\alpha\beta}~.
    $$
    The irreducible part of eq. (\ref{om1}) symmetric in $\dot{\alpha},\dot{\beta}$ and
    $\alpha,\beta$ results in
    $$
    \e_{\dot{\gamma}(\dot{\alpha}}
    \chi_{\dot{\beta})~\beta\alpha} = 0 \qquad \Rightarrow \qquad
    \chi_{\dot{\beta}~\beta\alpha} = 0~,
    $$
    while the part antisymmetric in $\dot{\alpha}\dot{\beta}$
    and $\alpha,\beta$ determines
    $\bar{\psi}_{0\dot{\beta}} $
    as the spinor derivative of the superfield $\Phi$
 \begin{equation}\label{phi0}
\bar{\psi}_{0\dot{\beta}} = {2i} \bar{D}_{\dot{\beta}}\Phi~.
\end{equation}
Thus we obtain eqs. \p{cfEA} and \p{cfEdA}.

    Substituting the results back into the above equation
    and analyzing the irreducible parts which remain we obtain
    eq. \p{cfw}.\footnote{The part with $[\dot{\alpha}\dot{\beta}], (\alpha\beta)$
    requires $\om_{\dot{\gamma}\alpha\beta}=0\,,$
    then the part
    with $(\dot{\alpha}\dot{\beta}), [\alpha\beta]$ gives the
    expression for
    $\bar{\om}_{\dot{\gamma}\dot{\beta}\dot{\alpha}}~.$}

To check the coefficient in the right hand side of \p{phi0} let us
substitute the expressions obtained (except for eq. \p{phi0}) into
eq. \p{om1}. We thus have
 \begin{equation}\label{om2}
     \e_{\alpha\beta}
     \bar{\om}_{\dot{\gamma} \dot{\beta}\dot{\alpha}} =
     -2 \e_{\alpha\beta}\e_{\dot{\alpha}\dot{\beta}} \bar{D}_{\dot{\gamma}}
\Phi - 2i
     \e_{\alpha\beta} \e_{\dot{\alpha}\dot{\gamma}}
     \bar\psi_{0\dot{\beta}}
\end{equation}
or, omitting $\e_{\alpha\beta}~,$
 \begin{equation}\label{om3}
     \bar{\om}_{\dot{\gamma} \dot{\beta}\dot{\alpha}} =
     -2 \e_{\dot{\alpha}\dot{\beta}} \bar{D}_{\dot{\gamma}} \Phi - 2i
     \e_{\dot{\alpha}\dot{\gamma}}
     \bar\psi_{0\dot{\beta}}\,.
\end{equation}
 Recall that
 $
 \bar{\om}_{\dot{\gamma} \dot{\beta}\dot{\alpha}} =
\bar{\om}_{\dot{\gamma} \dot{\alpha}\dot{\beta}}$ and hence $
\bar{\om}_{\dot{\gamma} \dot{\alpha}\dot{\beta}}
\e^{\dot{\alpha}\dot{\beta}}=0\,.$ Thus, contracting \p{om3} with
$\e^{\dot{\alpha}\dot{\beta}}$ one derives eq. \p{phi0}.

Now, substituting  \p{cfEA}, \p{cfEdA}, \p{cfw} and  \p{cf} into the fermionic
structure equations \p{DEA}, one finds from the analysis of the $d\theta^\beta
\wedge d\theta^\gamma$ component that
$$
 D_\beta(3\Phi + i W) =0~,
$$
or
 \begin{equation}\label{bphi}
      3\Phi + iW = 6\bar{\phi}~, \qquad
      D_\alpha\bar{\phi}=0~. \qquad
\end{equation}
The complex conjugate equation
\begin{equation}\label{phi}
 \bar{D}_{\dot{\beta}}(3\Phi - i W) =0 \quad \Rightarrow
\quad
3\Phi - iW = 6{\phi}~, \qquad
      \bar{D}_{\dot{\alpha}}{\phi}=0~,
\end{equation}
 follows from the
$\propto d\theta^\beta \wedge d\bar{\theta}^{\dot{\gamma}}$ component of \p{DEA}.

Thus  both $\Phi$ and $W$ are expressed through the
chiral and antichiral superfields
$\phi$ ($\bar{D}_{\dot{\alpha}}\phi =0$) and $\bar{\phi}$
 (${D}_{{\alpha}}\bar{\phi} =0$)
 \begin{equation}\label{PW}
      \Phi = \phi + \bar{\phi}~, \qquad
      iW = 3(\bar{\phi}-{\phi})~. \qquad
\end{equation}

The component of the basic form $\Pi^{\gamma\dot{\gamma}}\wedge
d\bar{\theta}^{\dot{\beta}}$ of eq. \p{DEA} produces the equation for the
chiral and antichiral superfields
 \begin{equation}\label{Eq}
     \bar{D}_{\dot{\alpha}}
      \bar{D}^{\dot{\alpha}} e^{ 4\bar{\phi}}
      = {4i \over R} e^{8\phi}\,.
\end{equation}
The left hand side of this equation contains only the antichiral
superfield while its right hand side contains only the chiral one.

The $\Pi^{\beta\dot{\beta}}\wedge d{\theta}^{{\gamma}}$ component of eq.
\p{DEA} specifies the form of the vector component $w_{\beta
\dot{\beta}\gamma}{}^\alpha$ of the spin connection
 \begin{eqnarray}\label{wa}
     w_{\beta \dot{\beta}\gamma}{}^\alpha=
4 \partial_{\beta \dot{\beta}}
\bar{\phi} \delta_\gamma{}^\alpha -
4 \partial_{\gamma \dot{\beta}}
\bar{\phi} \delta_\beta{}^\alpha - 2
\partial_{\beta \dot{\beta}}
{\phi} \delta_\gamma{}^\alpha - {8i} D_\beta {\phi}
\bar{D}_{\dot{\beta}}\bar{\phi} \delta_\gamma{}^\alpha + {8i} D_\gamma {\phi}
\bar{D}_{\dot{\beta}}\bar{\phi} \delta_\beta{}^\alpha \; .
\end{eqnarray}
Since by definition the spin connection is traceless,
$w_\alpha{}^\alpha \equiv 0$, one finds from \p{wa} that
 \begin{eqnarray}\label{EqW}
\partial_{\beta \dot{\beta}} ({\phi}- \bar{\phi})= -{2i}
 D_\beta {\phi} \bar{D}_{\dot{\beta}}\bar{\phi} \; ,
\end{eqnarray}
i.e., $\partial_{\beta \dot{\beta}}W= {6}
 D_\beta {\phi} \bar{D}_{\dot{\beta}}\bar{\phi}\,.$
Eq. \p{EqW} can be used to simplify the expression \p{wa} for
$w_{\beta \dot{\beta}\gamma}{}^\alpha$
 \begin{eqnarray}\label{wa1}
     w_{\beta \dot{\beta}\gamma}{}^\alpha=
2 \partial_{\beta \dot{\beta}} {\phi} \delta_\gamma{}^\alpha -4
\partial_{\gamma \dot{\beta}} {\phi} \delta_\beta{}^\alpha \; .
\end{eqnarray}
This expression can be also obtained directly from the $d\theta\wedge
d\bar{\theta} $ component of eq. \p{R}. As a consequence of eqs. \p{wa1},
\p{cfw} and \p{PW}, the `left chiral' spin connection is expressed through the
derivatives of the chiral field $\phi$ only
 \begin{equation}\label{cfw1}
    {w}_\beta^{~\alpha}=
    -2 d\theta^\alpha D_\beta \phi
    -2 d\theta_\beta D^\alpha \phi -
    2\Pi^{\alpha\dot{\gamma}} \partial_{\beta\dot{\gamma}} \phi +
2\Pi_{\beta\dot{\gamma}}\partial^{\alpha\dot{\gamma}} \phi\; .
   \end{equation}

To check the compatibility of eqs. \p{Eq} and \p{EqW} one should verify that
applying the derivative $\bar{D}_{\dot\beta}$ to eq. \p{EqW} produces the
equation which coincides with the one obtained by applying $D_{\beta}$ to eq.
\p{Eq}. This fermionic integrability condition can be written in the form
 \begin{eqnarray}\label{Eqf}
   \partial_{\beta \dot{\beta}} \bar{D}^{\dot{\beta}}\bar{\phi} =
-{2\over R} D_\beta \phi \; e^{8\phi- 4\bar{\phi}} - 4\partial_{\beta
\dot{\beta}}\bar \phi \bar{D}^{\dot{\beta}}\bar{\phi}\; .
\end{eqnarray}
However, the first fermionic component in the $\theta$-expansion of eq.
\p{EqW} is  more restrictive than the same component of eq.
\p{Eq}. Indeed, acting on eq. \p{EqW} by $\bar{D}_{\dot{\alpha}}$
(no contraction of the dotted indices is made in contrast to
\p{Eqf}), one finds
\begin{eqnarray}\label{Eqf2}
   \partial_{\beta \dot{\beta}} \bar{D}_{\dot{\alpha}}\bar{\phi} =
{i} \e_{\dot{\alpha}\dot{\beta}} \bar{D}\bar{D}\bar{\phi}\,
D_\beta \phi + \partial_{\beta\dot{\alpha}} \phi\,
\bar{D}_{\dot{\beta}}\bar{\phi}\; .
\end{eqnarray}
Eq. \p{Eqf2} also implies
\begin{eqnarray}\label{Eqb2}
   \partial_{\beta \dot{\beta}} \bar{D}\bar{D}\bar{\phi} =
4 \bar{D}\bar{D}\bar{\phi} \partial_{\beta \dot{\beta}}\phi \; .
\end{eqnarray}
At this step the check of consistency can be carried out by
verifying that the equation obtained by taking the derivative
$\partial_{\beta \dot{\beta}}$ of eq. \p{Eq} is satisfied
identically when eqs. \p{Eqb2} and \p{Eqf2} are taken into
account.

The $SO(1,3)$ invariant solutions of eqs. \p{Eq} and \p{EqW} specify the form
of the chiral superfield ${\phi}$ for the  $AdS_4$ coset superspace
${OSp(1|4;R)}\over{SO(1,3)}$ and thus define the superconformally flat
parametrization of the latter.

In a similar way one can demonstrate the superconformal flatness of the coset
superspace ${OSp(N|4;R)}\over{SO(1,3)\times SO(N)}$ for any $N$.

In Section 3 we shall describe a method which allows one to derive the explicit
form of the conformal factor $\Phi(x,\theta)$ and the phase factor
$W(x,\theta)$ for these supermanifolds without solving the Maurer--Cartan
equations.

\subsection{AdS$_2\times$S$^2$ superspaces}

Let us consider now a coset superspace ${{SU(1,1|2)}\over{SO(1,1)\times
SO(2)}}$ whose bosonic body is the space $AdS_2\times S^2\,.$ When the radii $R$
of $AdS_2$ and $S^2$ are equal, the metric of the bosonic space is conformally
flat.

We shall now show that the coset superspace ${{SU(1,1|2)}\over{SO(1,1)\times
SO(2)}}$ is however {\bf not} {\it super}conformally flat and that another
supercoset with the bosonic subspace $AdS_2\times S^2\,,$ namely
${OSp(4^*|2)}\over{SO(1,1)\times SO(2)\times SU(2)}\,$, is superconformally flat.
The supergroup ${OSp(4^*|2)}$ has eight fermionic generators, like $SU(1,1|2)$,
but a larger bosonic subgroup $SO(2,1)\times SO(3) \times SU(2)$ (see
Subsection 4.2 for details).\footnote{Note that the supergroup $SU(1,1|2)$ has
a group of outer automorphisms $SU(2)$ whose generators do not appear in the
anticommutator of the supercharges.}

The $ AdS_2\times S^2$ space is the Reissner--Nordstr\"om extreme black hole
solution of $N=2$, $D=4$ supergravity (as discussed, e.g. in
\cite{black,RK.98}). In \cite{RK.98} it was demonstrated that the corresponding
coset superspace ${{SU(1,1|2)}\over{SO(1,1)\times SO(2)}}$ is a solution of
$N=2$, $D=4$ superfield supergravity constraints. On this solution the $N=2$,
$D=4$ superspace coordinates $x^a=(x^{a'},x^{\hat a})$ $(a'=0,3;~\hat a=1,2)$,
$\theta^\alpha_i$ and $\bar\theta^{\dot\alpha\,i}$ ($i=1,2$) split into
$AdS_2\times S^2$ vectors and spinors, and  the supergravity constraints reduce
to the Maurer--Cartan structure equations of ${{SU(1,1|2)}\over{SO(1,1)\times
SO(2)}}\,.$

Instead of writing the Maurer--Cartan equations of
${{SU(1,1|2)}\over{SO(1,1)\times SO(2)}}$ we shall consider Maurer--Cartan
equations of the one-parameter set of supercosets
${D(2,1;\alpha)}\over{SO(1,1)\times SO(2)\times SU(2)}\,\,$, where $D(2,1;\alpha)$
is an exceptional supergroup with eight fermionic generators and the bosonic
subgroup $SO(2,1)\times SO(3) \times SU(2)\,,$ $\alpha$ being a numerical
parameter. When $\alpha=1\,,$ $D(2,1;\alpha)$ becomes isomorphic to
${OSp(4^*|2)}$ and, when $\alpha=-1$, it reduces to the semi--direct product of
${SU(1,1|2)}$ and the outer automorphism group $SU(2)$ (see \cite{sorba} and
Subsection 4.2 for details). Thus all the coset superspaces
${D(2,1;\alpha)}\over{SO(1,1)\times SO(2)\times SU(2)}$ contain $AdS_2\times
S^2$ as the bosonic body and possess the same number of
fermionic dimensions. The
supercosets ${OSp(4^*|2)}\over{SO(1,1)\times SO(2)\times SU(2)}$ and
${{SU(1,1|2)}\over{SO(1,1)\times SO(2)}}$ are recovered at $\alpha=1$ and
$\alpha=-1\,,$ respectively.

The torsion and the curvature constraints have the form
 \begin{eqnarray}\label{AdS2Ta=}
  & T^{{a'}} \equiv dE^{{a'}} - E^{{b'}}
\wedge  w_{{b'}}{}^{{a'}} & = - 2i\,
       E^{{\alpha}}_i  \wedge
       E^{\dot{\beta}i} \,
       \sigma^{{a'}}_{{\alpha}\dot{\beta}}\; ,\\
& T^{{\hat a}} \equiv dE^{{\hat a}} - E^{{\hat b}} \wedge  w_{{\hat
b}}{}^{{\hat a}} & = 2i\alpha \,
       E^{{\alpha}}_i  \wedge
       E^{\dot{\beta}i} \,
       \sigma^{{\hat a}}_{{\alpha}\dot{\beta}}\; ,
\end{eqnarray}
\begin{equation}
\label{AdS2Tal=}  T^{{\alpha}}_i \equiv dE^{{\alpha}}_i- E^{{\beta}}_i\wedge
w_{{\beta}}{}^{{\alpha}} - E^{{\alpha}}_j\wedge \omega_{j}{}^{i} =
{{2(1-i)}\over R} E^{{a}} \wedge \bar{E}^{\dot{\beta}}_j f^{\alpha\beta}
\sigma_{a\beta\dot{\beta}} \; ,
\end{equation}
\begin{equation}
\label{AdS2Tdal=} T^{\dot{\alpha}i} \equiv dE^{\dot{\alpha}i}-
E^{\dot{\beta}i}\wedge w_{\dot{\beta}}{}^{\dot{\alpha}}- \bar
E^{{\dot\alpha}}_j\wedge \omega_{j}{}^{i}  = - {{2(1+i)}\over R} E^{{a}} \wedge
E^{{\beta}}_j \epsilon^{ij} \sigma_{a\beta\dot{\beta}}
\bar{f}^{\dot{\alpha}\dot{\beta}} \; ,
\end{equation}
\begin{eqnarray}
\label{AdS2Rab=}
 & \!\!\!\!\!\!\!\!\!\!\!\! R^{{a'}{b'}} \equiv dw^{{a'}{b'}}-
w^{{a'}{c}}\wedge w_{{c}}{}^{{b'}} 
= - {i \over R}\left( E^{{\alpha}i}  \wedge
       E_{{\alpha}i}\,
 + \bar E_{\dot{\alpha}i}  \wedge
       \bar E^{\dot{\alpha}i} \right )
\epsilon^{a'b'}
 - {1\over 4 R^2}  E^{{d'}} \wedge E^{{c'}}
\epsilon_{c'd'}\epsilon^{a'b'},\\
\label{S2Rab=} & R^{{\hat a}{\hat b}} \equiv dw^{{\hat a}{\hat b}}-
w^{{\hat a}{c}}\wedge w_{{c}}{}^{{\hat b}} 
=  {\alpha\over R} \left(E^{{\alpha}i}  \wedge
       E_{{\alpha}i}\,
- \bar E_{\dot{\alpha}i}  \wedge
       \bar E^{\dot{\alpha}i}\right) \epsilon^{\hat a\hat b}
 + {1\over 4 R^2}  E^{{\hat d}} \wedge E^{{\hat c}}
\epsilon_{\hat c\hat d}\epsilon^{\hat a\hat b},
\end{eqnarray}
\begin{equation}
 \label{SU2R=}
R^{ij}=d\omega^{ij}-\omega^{ik}\wedge \omega_k^{~j}={{1+\alpha}\over{2R}}
\left((1-i)E^{i\alpha}\wedge E^{j\beta}f_{\alpha\beta}-(1+i)\bar
E^{i\dot\alpha}\wedge \bar E^{j\dot\beta}\bar f_{\dot\alpha\dot\beta}\right).
\end{equation}
Here
\begin{eqnarray}\label{Ef=EfEf}
E^{\underline{\alpha}} \equiv (E^{\a}_i\; , \bar{E}^{\dot{\a}i}) \,, \quad
\bar{E}^{\dot{\a}i}= (E^{\a}_i)^* \; ,
\end{eqnarray}
$w^{{a}{b}}$, $w_{\beta}^{~\alpha}={1\over 4}
w^{{a}{b}}(\sigma_{a}\tilde{\sigma}_b)_\gamma {}^{\alpha}$ and
$w_{\dot\beta}^{~\dot\alpha}=(w_{\beta}^{~\alpha})^*$ are $AdS_2\times S^2$
spin connections with the curvature $R^{ab}$, $\omega_{j}{}^{i}$ is the $SU(2)$
connection and $R^{ij}=R^{ji}$ is the corresponding curvature. Note that in the
case of ${{SU(1,1|2)}\over{SO(1,1)\times SO(2)}}\,,$ when $\alpha=-1\,,$ the
$SU(2)$ curvature $R^{ij}$ (\ref{SU2R=}) is zero and the $SU(2)$ connection can
be gauged away by an appropriate local $SU(2)$ transformation of $E^{\alpha i}$
and $\bar E^{\dot\alpha}_i\,.$

The symmetric spin tensors $f^{\alpha\beta}= f^{\beta\alpha}$,
$\bar{f}^{\dot{\alpha}\dot{\beta}}=
   \bar{f}^{\dot{\beta}\dot{\alpha}}= (f^{\alpha\beta})^*$ are related by
\begin{eqnarray}\label{Wss}
& f^{\alpha\beta}= {1\over 4} \epsilon^{\alpha\gamma}
(\sigma^a\tilde{\sigma}^b)_\gamma {}^{\alpha} f_{ab} \; , \qquad
\bar{f}^{\dot{\alpha}\dot{\beta}} =  {1\over 4}
\epsilon^{\dot{\beta}\dot{\gamma}}
(\tilde{\sigma}^a\sigma^b)^{\dot{\alpha}}{}_{\dot{\gamma}} f_{ab}
\end{eqnarray}
to the $SO(1,1)\times SO(2)$ invariant antisymmetric tensor
\begin{eqnarray}\label{2gframe} f_{ab} = \left(\matrix{
\e_{{a'}{b'}} & 0 \cr
 0 & \e_{\hat{a}\hat{b}} \cr}\right), \quad {a'}, {b'}=0,3\,;
\quad, \hat{a} , \hat{b} = 1, 2\,,
\end{eqnarray}
which in the case of the supercoset ${{SU(1,1|2)}\over{SO(1,1)\times SO(2)}}$
(i.e., when $\alpha=-1$) can be associated with the `vacuum' value of the field
strength $F_{ab}$ of the Abelian gauge field of the $N=2$, $D=4$ supergravity
multiplet. In the superspace the corresponding  constant closed form $F=dA$ is
\begin{equation}
F \epsilon^{ij} = {1\over 2} E^{{\alpha}i} \wedge
 E^{{\beta}j} \epsilon_{{\alpha}{\beta}}+
{1\over 2} \bar{E}^{\dot{\alpha}i} \wedge
 \bar{E}^{\dot{\beta}j} \epsilon_{\dot{\alpha}\dot{\beta}}
+ {1\over 2} E^a \wedge E^b f_{ab} \epsilon^{ij} \; , \quad dF\equiv 0 \; .
\end{equation}

One can verify that the constraints \p{AdS2Ta=}--\p{Ef=EfEf} admit a
superconformally flat solution only for $\alpha=1$, i.e. in the case of the
supercoset ${{OSp(4^*|2)}\over{SO(1,1)\times SO(2)\times SU(2)}}\,.$ The
superconformally flat form of the supervielbeins is
\begin{eqnarray}\label{2cfEa}
E^a = e^{\Phi + \bar{\Phi}} \Pi^bL_b^{~a}(x,\theta) \; , \qquad \Pi^a =dx^a - i
(d\theta^{\alpha}_i \sigma^a_{\alpha\dot{\beta}} \bar{\theta}^{\dot{\beta}i}-
\theta^{\alpha}_i
\sigma^a_{\alpha\dot{\beta}} d\bar{\theta}^{\dot{\beta}i})\, , \\
\label{2cfEal} E^{\alpha}_i = (1-i)\,e^{\bar{\Phi}} (d\theta^{\beta}_i +{i\over
2} \Pi^a \tilde{\sigma}_a^{\dot{\beta}\beta}
\bar{D}_{\dot{\beta}i}\bar{\Phi})L_\beta^{~\gamma}
f_{\gamma}^{~\alpha}\; , \\
\label{2cfEa2} \bar{E}^{\dot{\alpha}i} =(1+i)\,e^{{\Phi}} \;
(d\bar{\theta}^{\dot{\beta}i} - {i\over 2} \Pi^a
\tilde{\sigma}_a^{\dot{\beta}\beta} {D}^i_{\beta}{\Phi})\bar
L_{\dot\beta}^{~\dot\gamma}\bar f^{~\dot\alpha}_{\dot\gamma} \; ,
\end{eqnarray}
where
\begin{eqnarray}\label{bDPhi}
\bar{D}_{\dot{\beta}j}\Phi = 0\; , \qquad {D}^j_{\beta} \bar{\Phi}=0 \; ,
\qquad \bar{\Phi}=(\Phi )^*\,,
\end{eqnarray}
and $L_b^{~a}(x,\theta)$, $L_\beta^{~\alpha}(x,\theta)$ and $\bar
L_{\dot\beta}^{~\dot\alpha}(x,\theta)$ are some $SO(1,3)$ matrices in vector
and spinor representations, respectively.

The crucial point where the superconformally flat ansatz fails for the
supercoset ${{SU(1,1|2)}\over{SO(1,1)\times SO(2)}}$ is the spinor torsion
constraints \p{AdS2Tal=} and \p{AdS2Tdal=}, while these are compatible with the
superconformally flat ansatz \p{2cfEa}--\p{bDPhi} for
${OSp(4^*|2)}\over{SO(1,1)\times SO(2)\times SU(2)}$ and produce differential
equations for the Weyl and $U(1)$ factor, such as

\begin{eqnarray}\label{2.40a}
\bar{D}_{\dot{\gamma} i}\bar{D}^{\dot{\gamma}}_j \bar{\Phi} + 2
\bar{D}_{\dot{\gamma} i}\bar{\Phi}\; \bar{D}^{\dot{\gamma}}_j \bar{\Phi} =0 \;,
\qquad
\\ \label{EqbPhi2}
\bar{D}_{\dot{\alpha} i}\bar{D}_{\dot{\beta}}^i \bar{\Phi} - 2
\bar{D}_{\dot{\alpha} i}\bar{\Phi}\; \bar{D}_{\dot{\beta}}^i \bar{\Phi} = -
{8i\over R}e^{2\Phi} (\bar{L}\bar{f}\bar{L}^{-1})_{\dot{\alpha} \dot{\beta}}\;.
 \end{eqnarray}

Thus the $AdS_2\times S^2$ superspace which appears as a maximally
supersymmetric solution of $N=2$, $D=4$  supergravity is  not superconformal.
An intrinsic nature of this somewhat surprising feature, as well as the reason
why an alternative coset ${OSp(4^*|2)}\over{SO(1,1)\times SO(2)\times SU(2)}$
is superconformal are explained in detail in Subsection 4.2. They are traced to
the fact that $OSp(4^*|2)$ can be embedded as an appropriate subgroup into the
$N=2$, $D=4$ superconformal group, while $SU(1,1|2)$ cannot.
 In Subsection 4.2 we shall also derive an explicit
form of the conformal factor $\Phi(x,\theta)$ of \p{2cfEa} -- \p{2cfEa2}.

\subsection{ AdS$_3\times$S$^3$ superspaces}
The $N=2$, $D=6$ super Poincare group with {\bf 16} supercharges is  a
subsupergroup of the superconformal group in six dimensions
$U_{\alpha}(4|2;H)=OSp(8^*|4)$ which has {\bf 32} supercharges and
$O^*(8)\times USp(4)$ as the bosonic subgroup (see, e.g.
\cite{sconf,sorba,Claus:1998us} for a list of corresponding Lie superalgebras,
superconformal algebras and their subalgebras). A superspace whose bosonic
body is $AdS_3\times S^3$ is the supercoset ${{SU(1,1|2)\times
SU(1,1|2)}\over{SO(1,2)\times SO(3)}}\,,$ its isometry supergroup being the direct
product of two supergroups $SU(1,1|2)\,.$ The bosonic subgroup of this isometry
supergroup is $SU(1,1)\times SU(1,1)\times SU(2)\times SU(2)\,,$ and the
fermionic sector consists of {\bf 16} generators (supercharges). This
supercoset is a solution of $N=(2,0)$, $D=6$ supergravity. But, as in the case
of $AdS_2\times S^2$ considered above, the isometry supergroup
${SU(1,1|2)\times SU(1,1|2)}$ cannot be embedded as the appropriate subgroup
into the $D=6$ superconformal group $OSp(8^*|4)\,.$ Instead, the latter contains
as such a subgroup the supergroup ${{OSp(4^*|2)\times OSp(4^*|2)}}\,.$ Thus, it is the
coset superspace ${{OSp(4^*|2)\times OSp(4^*|2)}\over{SO(1,2)\times SO(3)\times
SU(2)\times SU(2)}}\,,$ also having $AdS_3\times S^3$ as the bosonic subspace and
{\bf 16} fermionic directions, which is superconformal.
 We have presented here only group-theoretical arguments, but by analogy
with the $AdS_2 \times S^2$ case one can also demonstrate this by analyzing the
relevant Maurer--Cartan equations or applying the `bottom--up' approach of
Sections 3, 4.

\subsection{AdS$_5\times$S$^5$ superspace}
The super  $AdS_5 \times S^5$ background is a maximally supersymmetric solution
of type IIB $D=10$ supergravity. If it were superconformally flat, then the
superconformally flat ansatz for $AdS_5 \times S^5$ would solve the type IIB
$D=10$ supergravity constraints (which are equivalent to the superfield
supergravity equations of motion).

We shall again see that the constraints which describe $AdS_5\times S^5$
superspace do not have a superconformally flat solution.\footnote{
Superconformally flat solutions of type IIB supergravity equations requires a
nontrivial axion--dilaton background which is zero in the $AdS_5 \times S^5$
superspace.} The reason is that the isometry supergroup $SU(2,2|4)$ of super
$AdS_5\times S_5$ is {\it not a sub--supergroup} of the superconformal group
$OSp(2|32;R)$ or $OSp(1|64;R)$ in ten dimensions.

The Maurer--Cartan structure equations for the relevant supercoset \linebreak $
{SU(2,2|4)\over SO(1,4)\times SO(5)} $ coincide with the type IIB supergravity
constraints \cite{IIB} restricted to the $AdS_5 \times S^5$ superbackground
\cite{RK.98} ($\underline{a}=0,1, \ldots , 9\, , \; \underline{\alpha}= 1,
\ldots 16\, , \; I=1,2$)

\begin{eqnarray}\label{AdSTa=}
  & T^{\underline{a}} \equiv dE^{\underline{a}} - E^{\underline{b}}
\wedge  w_{\underline{b}}{}^{\underline{a}}
& = - i\,
       E^{\underline{\alpha}I}  \wedge
       E^{\underline{\delta}I} \,
       \sigma^{\underline{a}}_{\underline{\alpha}\underline{\delta}}
\\ \nonumber && \equiv  - i\; \left(
       E^{\underline{\alpha}1}  \wedge
       E^{\underline{\delta}1}+       E^{\underline{\alpha}2}  \wedge
       E^{\underline{\delta}2}\right)\;
       \sigma^{\underline{a}}_{\underline{\alpha}\underline{\delta}}\; ,
\\
\label{AdSTal=}
& T^{\underline{\alpha}I} \equiv dE^{\underline{\alpha}I}-
E^{\underline{\beta}I}\wedge
w_{\underline{\beta}}{}^{\underline{\alpha}} & = {1\over R}
E^{\underline{a}} \wedge E^{\underline{\beta}J} \; \epsilon_{IJ}\;
f_{ \underline{a}\underline{b}_1\ldots \underline{b}_4}
(\sigma^{\underline{b}_1\ldots
\underline{b}_4})_{\underline{\beta}}
{}^{\underline{\alpha}}\; ,
\\
\label{AdSRab=}
& R^{\underline{a}\underline{b}} \equiv
dw^{\underline{a}\underline{b}}-
w^{\underline{a}\underline{c}}\wedge
w_{\underline{c}}{}^{\underline{b}} =  &
 -{1\over R^2}  E^{\underline{a}} \wedge E^{\underline{b}} -
{4i\over R^2} E^{\underline{\alpha}I}  \wedge
       E^{\underline{\delta}J} \epsilon_{IJ}
f^{
\underline{a}\underline{b}\underline{c}_1\underline{c}_2\underline{c}_3}
{\sigma}_{\underline{c}_1\underline{c}_2\underline{c}_3}
{}_{\underline{\alpha}\underline{\delta}}\; ,
\end{eqnarray}
where
\begin{eqnarray}
\label{wff=}
w_{\underline{\beta}}{}^{\underline{\alpha}}\equiv
{1\over 4} w^{\underline{a}\underline{b}} \sigma_{\underline{a}\underline{b}}
{}_{\underline{\beta}}{}^{\underline{\alpha}}\; ,
\end{eqnarray}
and $\sigma^{c_1...c_p}$ are antisymmetrized products of $p$ ten--dimensional
`Pauli' matrices.

The relative coefficients in  (\ref{AdSTa=})--(\ref{AdSRab=}) are
dictated by the Bianchi identities
\begin{eqnarray}
\label{BI1D} {\cal D} \bar{T}^{\underline{a}} = - \bar{E}^{\underline{b}}
\wedge  {R}_{\underline{b}}^{~\underline{a}}\; ,
 \\
\label{BI2D}
{\cal D} T^{{{\underline{\alpha}}}} =
- E^{{{\underline{\beta}}}} \wedge
R_{{{\underline{\beta}}}}{}^{{{\underline{\alpha}}}}\; ,
\\
\label{BI3D} {\cal D} {R}_{\underline{b}}^{~\underline{a}}=0\; .
\end{eqnarray}

Here $E^{\underline{a}}=(E^{\hat{a}}, E^i)\,,$ where $ {\hat{a}}=0,1,\ldots , 4$
is the vector index of the tangent space of $AdS_5\,,$ and $i=1,\ldots , 5$ is
the vector index of the tangent space of $S^5\,;$ $f_{ \underline{a}_1\ldots
\underline{a}_5}$ is a constant self--dual tensor of the following form
\begin{eqnarray}\label{gframe}
f_{\hat{a}_1\ldots\hat{a}_5}= c \varepsilon_{\hat{a}_1\ldots\hat{a}_5}\; ,
\quad \qquad f_{i_1\ldots i_5}= c \varepsilon_{i_1\ldots i_5}\; ,
\end{eqnarray}
with all other components vanishing. The constant $c$ is proportional to the
radius of $AdS_5$ or $S^5$ (these radii are equal).

We now check whether the superconformally flat ansatz for the
$AdS_5\times S^5$ superbackground
\begin{eqnarray}\label{cfEa}
E^{\underline{a}}= e^{2W} \Pi^{\underline{a}} \; , \qquad
\Pi^{\underline{a}}\equiv
d x^{\underline{a}} - i d\theta^{\underline{\alpha}I}\;
\sigma^{\underline{a}}_{\underline{\alpha}\underline{\beta}}
\theta^{\underline{\beta}I} \; ,
\end{eqnarray}
is compatible with eqs. \p{AdSTa=}--\p{AdSRab=}.

Substituting \p{cfEa} into (\ref{AdSTa=}) and analyzing  the lower dimensional
components of the latter
$$T_{\underline{\alpha}I\;
\underline{\beta}J}{}^a= - 2i \delta_{IJ}
\sigma^a_{\underline{\alpha}\underline{\beta}}~, \quad T_{\underline{\alpha}I\;
b}{}^a =0\,,
$$
one finds that the most general form of the fermionic
supervielbein consistent with both \p{cfEa} and  (\ref{AdSTa=}) is
\begin{eqnarray}\label{cfEal}
E^{\underline{\alpha}I}= e^{W} \left( d\Theta^{\underline{\alpha}J} - \, i \,
\Pi^{\underline{a}}
\tilde{\sigma}_{\underline{a}}^{\underline{\alpha}\underline{\beta}}
D_{\underline{\alpha}J}W \; \right)\; h_{J}^{\; I}\; ,
\end{eqnarray}
where
 \begin{eqnarray}\label{1hIJ=}
h_{J}^{\;I} = {1\over \sqrt{1+a^2}}
\left(\matrix{ 1
& -a \; \cr a & 1}\right)\equiv \left(\matrix{ \cos \alpha & - \,
\sin
\alpha \; \cr
 \sin \alpha &  \cos\alpha }\right)\;
\end{eqnarray}
is an $SO(2)$ matrix constructed from a real superfield
$a(x,\theta)$, with  $\alpha(x,\theta)= arccos{1\over \sqrt{1+a^2}}\,.$

However, as can be shown by a straightforward though tedious calculation, the
lowest dimensional component of eq. \p{AdSTal=}, $T_{\underline{\gamma}K\;
\underline{\beta}J}^{\; \underline{\alpha}I}=0 \;$,
 implies
 \begin{eqnarray}\label{DWDa}
D_{\underline{\alpha}I}W=0 \; , \qquad D_{\underline{\alpha}I}~
\alpha=0
\; ,
\end{eqnarray}
which requires the background superspace to be flat.

{We have already mentioned that the negative answer for the case of
$AdS_5\times S^5$ (as well as for the conventional $AdS_2\times S^2$ and
$AdS_3\times S^3$ superspaces) is explained by the following fact. Though the
type IIB $D=10$ super Poincar\'e group (with tensorial charges) is a subgroup of
 generalized simple superconformal groups $OSp(2|32;R)$ and $OSp(1|64;R)$ in ten
dimensions, the isometry supergroup $SU(2,2|4)$ of super $AdS_5\times S^5$ is
{\it not a sub--supergroup} of the superconformal group. At the same time, for
the background to possess a (super)conformal structure it is crucial that both
its (super)isometry group and (super)isometry of the flat (super)space are
subgroups of some encompassing (super)conformal group.

A simple and transparent way to see that $SU(2,2|4)$ cannot be a subgroup of
$OSp(2|32;R)$ and/or of $OSp(1|64;R)$ is as follows. A fundamental
representation of $SU(2,2|4)$ is a complex supertwistor \cite{ferber}
\begin{equation}\label{st}
Z^A=(\lambda^\alpha,\,\bar\mu_{\dot\alpha},\,\psi^i)~,
\end{equation}
where two Grassmann--even Weyl spinors $\lambda^\alpha$ and
$\bar\mu_{\dot\alpha}$ form a Dirac spinor with respect to the $D=4$ Lorentz
transformations, and  the complex Grassmann--odd components $\psi^i$
($i=1,2,3,4$)  are in the fundamental representation of the R--symmetry
subgroup $SU(4)$. Therefore the Grassmann--odd part of the supertwistor has
eight real components. (There are no representation with a less number of
Grassmann--odd components).

If the supertwistor obeys the following commutation relations
\begin{equation}\label{stc}
[Z^A,\,\bar Z_B]_\mp=\delta^A_B~,\quad [Z^A,\,Z^B]_\mp=0\,, \quad \bar
Z_B=(\mu^\beta,\,\bar\lambda_{\dot\beta},\,\bar\psi^i)
\end{equation}
($[,]_+$ stands for the anticommutator of the Grassmann--odd components), the
generators of \linebreak $SU(2,2|4)$ can be realized as the Hermitian bilinear
combination
\begin{equation}\label{suge}
M^A_{~B}=Z^A\bar Z_B\,.
\end{equation}

 Fundamental representations of $OSp(2|32;R)$ and $OSp(1|64;R)$ are
 real supertwistor (or supersingleton) representations
\begin{equation}\label{ss}
\Lambda_{\hat A}=(\lambda_{\hat\alpha},\chi^I)~,
\end{equation}
where now $\lambda_{\hat\alpha}$ is a Grassmann--even 32--component real spinor
or a 64--component real spinor and $\chi^I$ denotes the Grassmann--odd real
components of the supertwistor.  The $OSp(2|32;R)$ supertwistor contains two
real Grassmann--odd components, which form a vector of an $SO(2)$ subgroup of
$OSp(2|32;R)$, and the $OSp(1|64;R)$ supertwistor contains only one real
Grassmann--odd component.

If the components of  (\ref{ss}) satisfy the (anti)commutation
relations
\begin{equation}\label{ssc}
[\Lambda_{\hat A},\Lambda_{\hat B}]_\mp=C_{\hat A\hat B}\,,
\end{equation}
where $C_{\hat A\hat B}=(C_{\hat\alpha\hat\beta},\,\delta_{IJ})$ is an
$OSp$--invariant constant matrix, the generators of $OSp(2|32;R)$ and
$OSp(1|64;R)$ can be realized as the bilinear combination of the supertwistor
(\ref{ss}) components
\begin{equation}\label{ospg}
M_{\hat A\hat B}=\Lambda_{(\hat A}\Lambda_{\hat B]},
\end{equation}
where the matrix $M_{\hat A\hat B}$ is symmetric in the indices
$\hat\alpha,\,\hat\beta$ and antisymmetric in the indices $I, J$.

Now, if $SU(2,2|4)$ were a subgroup of $OSp(2|32;R)$ or $OSp(1|64;R)$, the
supersingleton representation (\ref{ss}), having been decomposed into irreps of
this subgroup, would contain the supertwistor representation (\ref{st}). But
this is obviously not the case since (\ref{st}) has eight real Grassmann--odd
components and (\ref{ss}) has only one or two. The above reasoning suggests
that the minimal simple supergroups which can simultaneously contain both
$OSp(2|32;R)$ and $SU(2,2|4)$ as sub--supergroups are $OSp(8|32;R)$ and
$SU(16,16|4)$. The type IIB $D=10$ super Poincar\'e group with a `central'
extension, as well as special superconformal transformations are contained in
$OSp(2|32;R)$, but the fermionic sector of $SU(2,2|4)$ is certainly not. Thus
even within these much larger supergroups\footnote{Note that these supergroups
have 256 Grassmann--odd generators, which is four times as much as the number
of spinor charges one assumes to be present in superconformal groups associated
with M--theory (i.e. 64 \cite{osp164,bars,west}).} we are not able to relate
(in the superconformal sense discussed in Sections 3 and 4) the flat D=10
superspace and the $AdS_5\times S^5$ superspace. This explains why the
superconformally flat ansatz is inconsistent with the $AdS_5\times S^5$
solution of the type IIB supergravity constraints. One may ask whether there
exists a different $AdS_5\times S^5$ supercoset, analogous to the ones of
$AdS_2\times S^2$ and $AdS_3\times S^3$ of Subsections 2.2 and 2.3, which is
superconformal. We have looked through the list of the real forms of the
classical Lie superalgebras (\cite{sorba}) and have not found another simple
superalgebra, different from $su(2,2|4)$, which could be the isometry of such a
supercoset. Thus it does not exist.\footnote{We would like to thank Paul Sorba
for the discussion of this point.}

  \setcounter{equation}{0}
\section{Superconformal flatness of AdS superspaces. `Bottom-up' approach}

\subsection{The idea of the method}
The general strategy of proving superconformal flatness for $AdS$ superspaces
and finding the relevant superconformal factors which we shall follow in this
Section applies to the cases when both the Poincar\'e supersymmetry and $AdS$
supersymmetry, with the equal number of supercharges and the equal number $D$
of translation generators in flat and AdS spaces, form two subgroups of the
superconformal group acting in a Minkowski superspace of the bosonic dimension
$D$. One starts from a coordinate realization of this superconformal group in
the Minkowski superspace \cite{ferber} and studies the transformation
properties of the relevant flat covariant differential of $x$ under this
realization. Generically, this differential is multiplied by a scalar weight
factor  and undergoes some induced (super coordinate dependent) Lorentz
transformation. Then one singles out the $AdS$ supersymmetry transformations
from the superconformal transformations as a linear combination of those of
Poincar\'e supersymmetry and those of the special conformal supersymmetry. As
the next step, one constructs, out of the original flat superspace coordinates,
the appropriate scalar density compensating the weight part of the
transformation of the flat covariant differential under the $AdS$
supersymmetry.
 The flat covariant differential of $x$ multiplied by
this factor undergoes only induced Lorentz transformations under the $AdS$
supersymmetry transformations and so it is the sought covariant differential of
the $AdS$ supersymmetry (the vector Cartan form $E^{a}$ of the previous
consideration). By construction, the corresponding interval is conformal to the
super Poincar\'e group  covariant interval, which proves the
superconformal flatness
of the $AdS$ superspace. In other words, one starts from the parametrization of
the $AdS$ superspace by the coordinates in which the super Poincar\'e subgroup
of the underlying superconformal group has the `canonical' manifest
realization. The precise relation of such a parametrization to the
parametrization where the $AdS$ subgroup has the `canonical' realization
(corresponding, e.g. to the exponential parametrization of the $AdS$ super coset
element) can be fairly complicated and we are not going to discuss this point
here. Also, we shall not give the precise form of the $AdS$--covariant
differentials of Grassmann coordinates (spinorial Cartan forms) in this
approach. Once we are aware of the superconformal factor, they can be restored,
up to a Lorentz rotation, by general formulas \p{superflat},
\p{cfEA},\p{cfEdA}, \p{2cfEal} and \p{2cfEa2}.

We shall start from the purely bosonic $D=4$ case and then consider the cases
of $N=1$ and $N=2$ $AdS$ supersymmetries in $D=4$. The case of general $N$ is
considered in Subsection 3.5 and the $AdS_2\times S^2$ case in Section 4.

\subsection{Toy example: bosonic AdS$_4$}
The special conformal transformations of the $D=4$ Minkowskian
coordinates $x^m \;(m=0, \ldots , 3)$ are as follows
\be
 \delta_K x^m =  2(b\cdot x) x^m- b^m x^2~, \label{1}
\ee where $b^m$ is a constant parameter. The covariant
differential $dx^m$ is transformed as follows
\be
 \delta_K dx^m = l^{[m n]}dx_n + 2 (b\cdot x) dx^m~, \qquad l^{[mn]} =
 2(x^mb^n - x^nb^m)~. \label{2}
\ee
The $AdS$ subgroup of $D=4$ conformal group is singled out via the
identification
 $$
 \delta_{AdS} = \delta_P + \delta_K\,,
 $$
 with
 $$
 b^m = m^2 a^m
 $$
and
\bea
 && \delta_{AdS} x^m = a^m + m^2 \left[  2 (a\cdot x) x^m -a^m x^2 \right]~,
 \label{3} \\
 &&
 \delta_{AdS} dx^m = \tilde{l}^{[m n]}dx_n + 2 m^2(a\cdot x) dx^m~, \qquad
 \tilde{l}^{[mn]} = 2m^2(a^nx^m - a^mx^n)~,\label{4}
\eea where $m$ is proportional to the inverse $AdS$ radius (the case of $-m^2$
would correspond to dS subgroup). Now we wish to construct the scalar factor
$f(x)$ which compensates for the weight factor $2 m^2(a\cdot x)$ in \p{4}. By
Lorentz covariance, it can depend only on $x^2 \equiv X$, $f(x) = f(X)$, and
should have the following transformation properties under \p{3}: \be
 \delta_{AdS} f(X) = -2m^2 (a\cdot x) f(X)~.
\ee
 From the last relation one obtains the simple equation for
$f(X)$:
 \be
 f' = -\frac{m^2}{1 +m^2 x^2}\; f~, \lb{5}
 \ee
which, up to a constant normalization factor, is solved by
 \be
 f(x^2) = {1 \over 1 +m^2 x^2}~.
 \ee
Hence, the object
 \be
 {\cal D}x^m = \frac{1}{1 + m^2 x^2} \;dx^m
 \ee
undergoes only induced Lorentz rotation under the $AdS$ translations and so it
is the $AdS$-covariant differential. Its square
 $$
 ds^2 = {\cal D}x^m{\cal D}x_m = \frac{1}{(1+m^2x^2)^2}\; dx^mdx_m
 $$
is the $AdS$ interval which in this parametrization is explicitly conformal to
the Minkowski interval $dx^mdx_m$. Just this parametrization of $AdS_4$ was
used e.g. in \cite{isor}.

\subsection{N=1, D=4 AdS superspace}
The $N=1$, $D=4$ superspace coordinates $(x^{\alpha\dot\alpha}, \theta^\alpha,
\bar\theta^{\dot\beta})$ are transformed in the following way under the special
supersymmetry of $N=1, D=4$ superconformal group (see e.g.
\cite{ferber}):\footnote{$\theta^\alpha = \epsilon^{\alpha\beta}\theta_\beta~,
 \;\;$ $\epsilon_{\alpha\beta}\epsilon^{\beta\gamma} = \delta^\gamma_\alpha~,
 \;\;$ $\theta^2 = \theta^\alpha\theta_\alpha~, \;\;\bar\theta^2 = \overline{\theta^2} =
 \bar\theta_{\dot\alpha}\bar\theta^{\dot\alpha}~.$}
 \bea &&\delta_\eta
 x^{\alpha\dot\alpha} = \eta^\lambda \theta^\alpha x_\lambda^{\;\;\dot\alpha}
 +{i\over 2} \eta^\alpha \bar\theta^{\dot\alpha} \theta^2 -\bar\eta^{\dot\rho}
 \bar\theta^{\dot\alpha} x^\alpha_{\;\;\dot\rho} + {i\over 2}
 \bar\eta^{\dot\alpha} \theta^{\alpha} \bar\theta^2~, \nn   &&
  \delta_\eta
 \theta^\rho = \eta^\rho \theta^2 - i\bar\eta^{\dot\rho}x_{\;\;\dot\rho}^\rho -
 \bar\eta^{\dot\rho} \bar\theta_{\dot\rho} \theta^\rho~,  \nn
 &&\delta_\eta \bar\theta^{\dot\rho} = \bar\eta^{\dot\rho} \bar\theta^2 +
 i\eta^{\rho}x_{\rho}^{\;\;\dot\rho} + \eta^\rho
 \theta_\rho \bar\theta^{\dot\rho}~. \lb{N1eta}
 \eea
It is enough to consider transformations with odd parameters since
all the remaining superconformal transformations are contained in
the closure of this conformal supersymmetry and the standard
Poincar\'e supersymmetry:
 \bea
 \delta_\epsilon\theta^\alpha = \epsilon^\alpha~,
 \;\;\delta_\epsilon\bar\theta^{\dot\alpha} = \bar\epsilon^{\dot\alpha}~, \;\;
 \delta_\epsilon x^{\rho\dot\rho} = - i(\epsilon^\rho\bar\theta^{\dot\rho} +
 \bar\epsilon^{\dot\rho} \theta^\rho)~. \lb{N1eps}
 \eea
The flat covariant differential of $x~,$
 \be
 \Pi^{\alpha\dot\alpha} = dx^{\alpha\dot\alpha} - i(d\theta^\alpha
 \bar\theta^{\dot\alpha} +d\bar\theta^{\dot\alpha} \theta^\alpha)~,
 \ee
is evidently invariant under \p{N1eps} and is transformed in the
following way under \p{N1eta}
 \be
 \delta \Pi^{\alpha\dot\alpha} = 2 \eta^{(\alpha}\theta^{\lambda)}
 \Pi_{\lambda}^{\;\;\dot\alpha} - 2
 \bar\eta^{(\dot\alpha}\bar\theta^{\dot\lambda)}
 \Pi_{\;\;\dot\lambda}^{\alpha} +\left( \eta\theta +
 \bar\eta\bar\theta\right) \Pi^{\alpha\dot\alpha}~.
 \label{N1etadx}
 \ee
The first two terms are the induced Lorentz rotation, while the last term is
the weight transformation. Now we single out the $AdS_4$ supertranslations as
the linear combination of \p{N1eps} and \p{N1eta} with
 $$
 \eta^\alpha  \rightarrow m \epsilon^\alpha~, \quad \bar\eta^{\dot\alpha}
 \rightarrow m \bar\epsilon^{\dot\alpha}~,
 $$
where $m$ is an arbitrary real parameter of the contraction to the
Poincar\'e supersymmetry. Like in the $N=0$ case we need to find
the real scalar factor
 $$
 A(x, \theta, \bar\theta) = \bar A~,
 $$
such that under the $AdS$ supersymmetry it has the following transformation
property
 \be
 \delta_{AdS} A = -m \left( \epsilon\theta +\bar\epsilon\bar\theta \right)A~.
 \lb{Atran} \ee
To preserve Lorentz covariance, it can depend only on the following invariants
 $$
 x^2 \equiv x^{\alpha\dot\alpha}x_{\alpha\dot\alpha}~, \quad \theta^2~, \quad
 \bar\theta^2~, \quad
 x^{\alpha\dot\alpha}\theta_\alpha\bar\theta_{\dot\alpha}~.
 $$
Fortunately, the terms proportional to the last nilpotent
invariant vanish (it can be directly checked) and $A$ has the
following $\theta$ expansion
 \be
 A = a(x^2) + c(x^2) \theta^2 +\bar c(x^2) \bar\theta^2 +
 d(x^2)\theta^2\bar\theta^2~. \lb{Aexpan}
 \ee
 From the condition \p{Atran} we get the set of equations for the
coefficients in  \p{Aexpan}
 \bea
 && a' - m c = a' - m\bar c = 0 \quad \rightarrow \quad c= \bar c~, \nn
 && mx^2 a'  + ma + 2c = 0~, \nn
 && ma' -2c' + 2m d = 0~, \nn
 && mx^2 c' + 3mc + 2 d =0~. \lb{N1set}
 \eea
The first two equations yield
 \be
 a' \left(1 +{m^2\over 2} x^2 \right) = -{m^2\over 2} a~,
 \ee
i.e., the same equation as \p{5} (up to rescaling of $m$). Thus,
 \be
 a = \frac{\beta}{1+{m^2\over 2}x^2}~,
 \ee
where $\beta$ is an arbitrary integration constant. The rest of equations
allows one to uniquely restore other functions, and the resulting expression
for $A$ proves to be as follows
 \bea
 A =  \frac{\beta}{(1+{m^2\over 2}x^2)}\left[1 - {m\over
 2}\frac{1}{(1+{m^2\over 2}x^2)}(\,\theta^2 + \bar\theta^2\,)
 +{m^2\over 4}\frac{3 + {m^2\over 2}x^2}{(1+{m^2\over
 2}x^2)^2}\,\theta^2\bar\theta^2 \right].\lb{N1A}
 \eea
Hereafter we choose $\beta = 1$.

One can obtain more elegant expression for $A$ in terms of
 $$
 x^{\alpha\dot\alpha}_L = x^{\alpha\dot\alpha} + i\theta^\alpha
 \bar\theta^{\dot\alpha}~, \qquad x^{\alpha\dot\alpha}_R= x^{\alpha\dot\alpha}
 -i\theta^\alpha \bar\theta^{\dot\alpha}~.
 $$
The factor $A$ has the following product structure
 \bea
 && A = f(x_L, \theta)\bar f(x_R, \bar\theta)~, \quad \bar f = \overline{(f)}
 \lb{N1prod} \\
 && f(x_L, \theta) = \frac{1}{\sqrt{1 + {m^2\over 2}x^2_L}}
 \left[1 - {m\over 2}\,\frac{1}{(1+{m^2\over 2}x_L^2)}\,\theta^2 \right].
 \eea
The $N=1$ covariant A$dS$ $x$-differential is now defined as
 \be
 {\cal D} x^{\alpha\dot\alpha} = f(\zeta_L)\bar f(\zeta_R)
 \Pi^{\alpha\dot\alpha} \qquad \left(\zeta_{L} =(x_L, \theta), \zeta_R
 = \overline{(\zeta_L )}\right)~.
 \ee
It undergoes only an induced Lorentz transformation under the $AdS_4$
supertranslations (and the appropriate one under the  $AdS_4$ translations).
Thus the $N=1$ super $AdS_4$ invariant interval is given by
 \be
 ds^2 = {\cal D} x^{\alpha\dot\alpha}{\cal D} x_{\alpha\dot\alpha}
 = f^2\bar f^2 \Pi^{\alpha\dot\alpha}\Pi_{\alpha\dot\alpha}~,
 \ee
and it is manifestly conformal to the Poincar\'e SUSY invariant
interval, in full correspondence with the derivation based on the
Maurer-Cartan equations. The above chiral factors coincide with
the factors entering the superfield Weyl transformations defined
in \cite{isor}, \cite{isor1}.

To establish a contact with the consideration in Section 2 , one can check that
the chiral multipliers $f(x_L, \theta), \bar f(x_R, \bar\theta)$ of the
superconformal factor obey the equation
$$
\bar D_{\dot\alpha} \bar D^{\dot \alpha} \bar f^2 = 4m f^4
$$
and its conjugate (in the left-chiral parametrization $D_\alpha =
\frac{\partial}{\partial \theta^\alpha} +
2i\bar\theta^{\dot\beta}\partial^L_{\alpha\dot\beta}~, \;\bar D_{\dot\alpha} =
-\frac{\partial}{\partial \bar\theta^{\dot\alpha}}$). One can reduce this
equation just to \p{Eq} by making a proper complex rescaling of $f, \bar f$.
Also, it is straightforward to check the relation
$$
f\,\partial^L_{\alpha\dot\beta}\bar f
-\bar f\,\partial^L_{\alpha\dot\beta} f =
iD_\alpha f\,\bar D_{\dot\beta}\bar f~,
$$
which amounts to eq. \p{EqW}. Thus the `bottom-up' method directly
yields the precise form of the required
particular Lorentz invariant solution of \p{Eq}, \p{EqW}. Note that the
linear equations
\p{N1set} (or their analogs for $f$ or $\bar f$) are obviously easier to solve
than the nonlinear equations \p{Eq} and \p{EqW}.

\subsection{N=2, D=4 AdS superspace}
In this case it is more convenient to work directly in the chiral
basis
 $$
 x^{\alpha\dot\alpha}_L = x^{\alpha\dot\alpha} +2i
 \theta^\alpha_i\bar\theta^{\dot\alpha i}, \quad \theta^\alpha_i,
 \quad \bar\theta^{\dot\alpha i}\,.
 $$
Under the special conformal and Poincar\'e $N=2$ supersymmetries
these coordinates transform as follows
 \bea
 &&\delta_\eta x_L^{\alpha\dot\alpha}= -4i x_L^{\beta\dot\alpha}\eta^i_\beta
 \theta^\alpha_i~, \nn
 && \delta_\eta \theta^{\alpha}_{i} = -2i \eta_{\beta i}
 \theta^{(\beta}_k\theta^{\alpha) k}  - 2i \eta^{\alpha}_k
 \theta^{\beta}_{(i}\theta^{k)}_\beta + \bar\eta_{\dot\beta i}
 x_L^{\dot\beta\alpha}~,
 \quad \delta_\eta \bar\theta^{\dot\alpha i} = \overline{(\delta_\eta
 \theta^\alpha_i)}~, \lb{N2eta} \\
 && \delta_\epsilon x_L^{\alpha\dot\alpha} = -4i \bar\epsilon^{\dot\alpha
 i}\theta^\alpha_i~,
 \quad \delta_\epsilon \theta^\alpha_i = \epsilon^\alpha_i~, \quad
 \delta_\epsilon \bar\theta^{\dot\alpha i}
 = \bar\epsilon^{\dot\alpha i}~, \lb{N2eps} \\
 && \bar\epsilon^{\dot\alpha i} = \overline{(\epsilon^\alpha_i)}~, \quad
 \bar\eta^{\dot\alpha}_{i} = \overline{(\eta^{\alpha i})}~. \nonumber
 \eea
Up to induced Lorentz rotations, the flat covariant differential
 \be
 \Pi^{\alpha\dot\alpha} = d x^{\alpha\dot\alpha} - 2i
 \left(d\theta^\alpha_i\bar\theta^{\dot\alpha i} + d\bar\theta^{\dot\alpha
 i}\theta^{\alpha}_i \right)
 \lb{N2int}
 \ee
is transformed as
 \be
 \delta_\eta \Pi^{\alpha\dot\alpha} = 2i(\eta^{\rho i}\theta_{\rho i})\;
 \Pi^{\alpha\dot\alpha}
 + \mbox{c.c.}~. \label{N2scint}
 \ee

A new feature compared to the previous case is that the $N=2$ $AdS_4$
supersymmetry is extracted via the following identification
 \be
 \delta_{AdS} = \delta_\epsilon + \delta_\eta~, \quad \eta^i_\rho =
 c^{(ik)}\epsilon_{\rho k}~, \;\;  \bar\eta^i_{\dot\rho} =
 -c^{(ik)}\bar\epsilon_{\dot\rho k}~, \;\;\overline{(c^{(ik)})} =
 \epsilon_{il}\epsilon_{ks}c^{(ls)}~,
 \ee
where a constant real vector $c^{(ik)}$ breaks the internal symmetry subgroup
$SU(2)$ of the $N=2$ superconformal group $SU(2,2|2)$ down to $SO(2)$ which is
the internal symmetry subgroup of the $N=2$ $AdS_4$ supergroup $OSp(2|4;R)$.

Thus, under the $N=2$ $AdS$ supersymmetry the flat supercovariant differential
$\Pi^{\alpha\dot\alpha}$ is transformed  (modulo induced Lorentz rotations) as
 \be
 \delta_{AdS}\Pi^{\alpha\dot\alpha} =
 -2i c^{(ik)}\epsilon_{\rho k}\theta^{\rho}_{i}\; \Pi^{\alpha\dot\alpha}
 + \mbox{c.c.}~. \lb{N2inttran}
 \ee
In order to find the compensating scalar factor, we assume that, similar to the
previous case, it is factorized into the product of chiral and anti-chiral
(conjugate) factors. So we need to find a complex factor $B(x_L, \theta)$, such
that it transforms under the $N=2$ $AdS$ supertranslations as follows
 \be
 \delta_{AdS} B =
 2i c^{(ik)}\epsilon_{\rho k}\theta^{\rho}_{i}\, B \lb{N2transfB}\,.
 \ee

The general $\theta$ expansion of $B$ consistent with the Lorentz
and $SO(2)$ invariances reads
 \be
 B = b_0(y) + \theta^{\alpha (i}\theta^{k)}_\alpha c_{ik} b_1(y) +
 \theta^{\alpha (i}\theta^{k)}_\alpha\theta^{\beta}_{(i}\theta_{\beta k)}
 b_2(y)~, \quad y \equiv x_L^2~. \lb{N2expan}
 \ee

As in the previous example, the transformation rule
\p{N2transfB} amounts to a set of the first-order linear
differential equations for the coefficients in \p{N2expan}
 \bea
 &&(\mbox{a})\quad 2i\,yb_0' +ib_0 + b_1 =0~, \nn
 &&(\mbox{b}) \quad 8ib_0' - c^2 b_1 = 0~, \nn
 &&(\mbox{c}) \quad 4ib_1' - 3 b_2 = 0~, \nn
 &&(\mbox{d}) \quad ic^2 b_1 +{2i\over 3} c^2 y b_1' + 2b_2 = 0~, \lb{N2eqs}
 \eea
where $c^2 = c^{ik}c_{ik}~.$ The following identities are useful while
extracting the independent structures in \p{N2transfB} in the course of
deriving \p{N2eqs}:
 $$
 c^{mn}\epsilon^\alpha_n\theta_{\alpha m} \phi = {1\over 3} c^2
 \epsilon^\alpha_n\theta_{\alpha m}\phi^{(mn)}~,
 \quad \phi \equiv c^{ik}\theta^{\beta}_i\theta_{\beta k}~, \;\;
 \phi^{(mn)} \equiv \theta^{\beta m}\theta_{\beta}^{n}~.
 $$
 $$
 \phi \phi = {1\over 3}c^2 \phi^{(ik)}\phi_{(ik)}~.
 $$

Eqs. (\ref{N2eqs}a), (\ref{N2eqs}b), (\ref{N2eqs}c) allow one to find $b_0$,
$b_1$ and
 $b_2$:
 \be
 b_0 = \frac{1}{\sqrt{1+{1\over 4}c^2y}}~, \quad b_1 = -i \frac{1}{(1+{1\over
 4}c^2y)^{3/2}}~,
 \quad b_2 = -{1\over 2}\frac{c^2}{(1+{1\over 4}c^2y)^{5/2}}~.
 \ee
Eq. (\ref{N2eqs}d) is then satisfied identically.

The final answer for $B$ is
 \bea
 B &=& \frac{1}{\sqrt{1+{1\over 4}c^2x^2_L}}\left[ 1 - i\phi\,\frac{1}{(1+{1\over
 4}c^2x^2_L)}
 -{1\over 2} \phi^{ik}\phi_{ik}
 \frac{c^2}{(1+{1\over 4}c^2x_L^2)^{2}} \right] \nn
 &=& \frac{1}{\sqrt{1+{1\over 4}c^2x^2_L}}\left[ 1 - i\phi\,\frac{1}{(1+{1\over
 4}c^2x^2_L)}
 -{3\over 2} \phi^{2}
 \frac{1}{(1+{1\over 4}c^2x_L^2)^{2}} \right].
 \eea
The $N=2$ $AdS$--covariant differential reads
 \be
 {\cal D}x^{\alpha\dot\alpha} = B\bar B\; \Pi^{\alpha\dot\alpha} \lb{N2AdSint}
 \ee
and the invariant interval is
 \be
 ds^2 = B^2\bar B^2\; \Pi^{\alpha\dot\alpha}
 \Pi_{\alpha\dot\alpha}~. \lb{N2AdSdist}
 \ee

\subsection{The case of arbitrary N}

The above construction works for any $N$ in $D=4$. In the generic case of
arbitrary $N$ the special conformal and Poincar\'e supersymmetry
transformations (both embedded into $SU(2,2|N)$) of the coordinates of the
$N$-extended Poincar\'e superspace in the left-chiral parametrization are given
by \cite{ferber}
 \bea
 &&\delta_\eta x_L^{\alpha\dot\alpha}= -4i x_L^{\beta\dot\alpha}\eta^i_\beta
 \theta^\alpha_i~, \nn
 && \delta_\eta \theta^{\alpha}_{i} = 4i \eta_{\beta}^k
 \theta^{\beta}_i\theta^{\alpha}_k + \bar\eta_{\dot\beta i} x_L^{\dot\beta\alpha}~,
 \quad \delta_\eta \bar\theta^{\dot\alpha i} = \overline{(\delta_\eta
 \theta^\alpha_i)}~, \lb{Neta} \\
 && \delta_\epsilon x_L^{\alpha\dot\alpha} = -4i \bar\epsilon^{\dot\alpha
 i}\theta^\alpha_i~,
 \quad \delta_\epsilon \theta^\alpha_i = \epsilon^\alpha_i~, \quad
 \delta_\epsilon \bar\theta^{\dot\alpha i}
 = \bar\epsilon^{\dot\alpha i}~, \lb{Neps} \\
 && \bar\epsilon^{\dot\alpha i} = \overline{(\epsilon^\alpha_i)}~, \quad
 \bar\eta^{\dot\alpha}_{i} = \overline{(\eta^{\alpha i})}~. \nonumber
 \eea
Now the indices $i,k, \ldots $ run from $1$ to $N$ (they correspond to
the fundamental representation of $SU(N)$) and the objects with
the upper-case and lower-case indices are no longer equivalent to
each other (as distinct from the special $N=2$ case, no analog of
$\epsilon_{ik}$ exists). The super Poincar\'e invariant
differential is still given by eq. \p{N2int}.

The super $ {AdS}_4$ subgroup $OSp(N|4;R)$ is singled out from $SU(2,2|N)$ via
the following identification
 $$
 \delta_{AdS} = \delta_\epsilon + \delta_\eta~, \qquad \eta^i_\alpha =
 C^{(ik)}\epsilon_{\alpha k}~,
 $$
where $C^{(ik)}$ is a constant symmetric tensor which breaks $SU(N)$ down to
$SO(N)$. Using the $SU(N)/SO(N)$ freedom, one can always bring $C^{(ik)}$ to
the diagonal form, $C^{(ik)} = {const_i}\times \delta^{ik}$. We shall use the
normalization
 \be
 C^{(ik)}\bar C_{(ij)} = {1\over N}\delta^k_j c^2~.
 \ee

It is easy to check that in the generic case the differential
$\Pi^{\alpha\dot\alpha}$ is transformed under the $\eta$
transformations, modulo induced Lorentz rotations, by the same law
\p{N2inttran}, now with $i=1, \ldots , N$. Thus we should now try to
construct general left-chiral superfunction $B(x_L, \theta)$ with
the transformation law
 \be
 \delta_{AdS} B =
 2i C^{(ik)}\epsilon_{\rho k}\theta^{\rho}_{i}\, B \lb{NtransfB}\,.
 \ee

Let us take the following general ansatz for $B$:
 \be
 B = b_0(y) + \phi b_1(y) + \phi^2 b_2(y) + \cdots + \phi^n b_n(y) + \cdots
 +\phi^N b_N(y)~, \lb{NB}
 \ee
where
 $$
 \phi = C^{(ik)}\theta^\alpha_i\theta_{\alpha k}~, \quad y = x_L^2~.
 $$
The latter invariants have the following transformation properties under the
$AdS$ supertranslations:
 \bea
 && \delta_{AdS} \phi = 4i C^{(ik)}\epsilon^\alpha_i\theta_{\alpha k}\phi +
 {2\over N} c^2 \bar \epsilon^i_{\dot\alpha}\theta_{\alpha i} x_L^{\dot\alpha\alpha}
 + 2 C^{(ik)}\epsilon^\alpha_i\theta_{\alpha k}~, \nn
 && \delta_{AdS} x^2_L = -8i \bar\epsilon^{\dot\alpha i}\theta^\alpha_i
 x_{L \alpha\dot\alpha} + 4i x_L^2 C^{(ik)}\epsilon^\alpha_i\theta_{\alpha k}~.
 \lb{}
 \eea

A thorough inspection of the conditions imposed on the coefficients in the
expansion \p{NB} by requiring $B$ to transform according to \p{NtransfB} shows
that one always gets two independent equations for each two consecutive
coefficients, e.g. for $b_0$ and $b_1$, $b_1$ and $b_2$, $b_2$ and $b_3$, etc.
For each pair these equations (obtained by putting to zero the coefficients of
$\epsilon$ and $ \bar\epsilon$ in the appropriate variations) form a closed set
and determine the relevant coefficients up to integration constants, being
different for different  pairs. Since each coefficient (except for $b_0$)
appears within two adjacent pairs, there arise relations between these
integration constants which finally fix $B$ up to an overall constant which we
choose, as in the previous particular cases, equal to 1.

Leaving the detailed calculations for an inquisitive reader (they
are much like to those in the previous subsections), let us give
the final surprisingly simple answer for $B(x_L, \theta)$:
\be\label{B}
 B = \frac{1}{\sqrt{1 +{c^2\over 2N}x_L^2 + 2i\phi}}~.
\ee
 Its expansion in powers of $\phi=C^{(ik)}\theta^\alpha_i\theta_{\alpha k}$
 automatically terminates at $\phi^N$ due
to the evident Grassmann property $\phi^{N+1} = 0$. It can be checked that at
$N=1, 2$ eq. (\ref{B}) reproduces the  chiral superconformal factors obtained
in previous subsections.

The $AdS$--covariant differential ${\cal D}x^{\alpha\dot\alpha}$ and the
invariant interval are defined in the same way as above, i.e. by eqs.
\p{N2AdSint} and \p{N2AdSdist}.

\setcounter{equation}{0}

\section{AdS$_2\times$S$^2$ superspace}
\subsection{Bosonic case}

The manifold ${AdS}_2\times S^2$ is a coset ${{SO(2,1)}\over{SO(1,1)}}\times
{{SO(3)}\over{SO(2)}}\,.$ The mutually commuting sets of the $SO(2,1)$ and
$SO(3)$ generators are singled out in the set of generators of the $M_4$
conformal group $SO(2,4)$ in the following way (in the notation of Subsect.
3.2)
 \bea
 && SO(2,1): \quad \left( P_{a'} + m^2 K_{a'}~, \;M_{{a'}{b'}} \right)~, \nn
 && SO(3): \quad \left( P_{\hat a} - m^2 K_{\hat a}~, \;M_{\hat a\hat b} \right)~,\quad
 {a'} = 0, 3~,
 \eta^{{a'}{b'}} = \mbox{diag}(1, -1);\quad  \hat a =
 1,2~.\label{so12so3gen}
 \eea

 This choice is unique up to an $SO(3)$ rotation of the
space-like indices $1,2,3$ and it is most convenient for supersymmetrizing. It
is easy to check that these two sets of the $SO(2,4)$ generators indeed commute
with each other; the first one comes from the $ {AdS}_4$ subgroup $SO(2,3)$
while the second one from the $ {dS}_4$ subgroup $SO(1,4)$. It is important to
note that the commutativity is possible only with the same contraction
parameter $m^2$ in both the $AdS_2$ and $S^2$ translation generators, which
amounts to the equal radii of the $AdS_2$ and $S^2$ from the geometric point of
view.

On the original Minkowskian set of coordinates $x^m \equiv (x^{a'}~, y^{\hat
a})$ the translation generators $P_{a'} + m^2 K_{a'}$ and $P_{\hat a} - m^2
K_{\hat a}$ of these two subgroups act in the following way (we denote the
relevant variations by the indices $1$ and $2$, respectively, and the
parameters by $a^{a'}$ and $c^{\hat a}$)
 \bea
 && \delta_1 x^{a'} = a^{a'} \left[1 - m^2 \left(x^2 - y^2\right) \right] + 2 m^2
 (a\cdot x)
 x^{a'}~,\quad \delta_1 y^{\hat a} = 2 m^2 (a\cdot x) y^{\hat a}~, \nn
 && \delta_2 y^{\hat a} = c^{\hat a} \left[1 + m^2 \left(x^2 - y^2\right) \right]
 + 2 m^2 (c\cdot y)
 y^{\hat a}~,\quad \delta_2 x^{a'} = 2 m^2 (c\cdot y) x^{a'}~. \label{tranprod}
 \eea
Hereafter,  $x^2 \equiv (x^0)^2 - (x^3)^2~, y^2 \equiv (y^1)^2
+(y^2)^2$ and analogously for $(a\cdot x)$ and $(c\cdot y)$.

We see that $x^{a'}$ and $y^{\hat a}$ do not form closed sets under these two
commuting groups.  The irreducible sets $z^{a'}$ and $t^{\hat a}$ can be
defined by the following invertible  relations
 \bea
 && z^{a'} = \frac{2}{A + \sqrt{A^2 + 4m^2 x^2}}\,x^{a'}~, \quad A = 1
 -m^2\left(x^2 - y^2\right)~, \nn
 && t^{\hat a} = \frac{2}{B + \sqrt{B^2 + 4m^2 y^2}}\,y^{\hat a}~, \quad B = 1
 +m^2\left(x^2 - y^2\right)~, \label{ztxy} \\
 && x^{a'} = \frac{1 + m^2 t^2}{1 - m^2 z^2t^2} z^{a'}~, \quad
 y^{\hat a} = \frac{1 + m^2 z^2}{1 - m^2 z^2t^2} t^{\hat a}~, \label{xyzt}
 \eea
 \bea
 && \delta_1 z^{a'} = a^{a'} \left(1 - m^2 z^2\right) + 2 m^2 (a\cdot z)
 z^{a'}~, \quad \delta_1 t^{\hat a} = 0~,\nn
 && \delta_2 t^{\hat a} = c^{\hat a} \left(1 - m^2 t^2\right) + 2 m^2 (c\cdot t)
 t^{\hat a}~, \quad \delta_2 z^{a'} = 0~.
 \eea
The covariant differentials of $z^{a'}$ and $t^{\hat a}$ can now be found by
the method of Subsect. 3.2:
 \be
 {\cal D}z^{a'} = {1\over 1 + m^2 z^2}\,dz^{a'}~, \quad {\cal D}t^{\hat a} =
 {1\over 1 + m^2 t^2}\,dt^{\hat a}~. \label{fullycov}
 \ee
They merely undergo  induced $SO(1,1)$ and $SO(2)$ rotations in
their indices, so
 $$
 {\cal D}z^{a'} {\cal D}z_{a'} = {1\over (1 + m^2 z^2)^2} dz^{a'} d
 z_{a'}~,
 $$
 and
 $$
 {\cal D}t^{\hat a}{\cal D}t^{\hat a} = {1\over (1 + m^2 t^2)^2} dt^{\hat a} d t^{\hat
 a}
 $$
are the corresponding invariant intervals.

Now, using the relations
 $$
 1 + m^2 z^2 = \frac{2\sqrt{A^2 + 4 m^2 x^2}}{A +\sqrt{A^2 + 4 m^2 x^2}}~, \;
 1 + m^2 t^2 = \frac{2\sqrt{B^2 + 4 m^2 y^2}}{B +\sqrt{B^2 + 4 m^2 y^2}}~,
 $$
and
 \be
 A^2 + 4 m^2 x^2 = B^2 + 4 m^2 y^2 \equiv F^2 (x^2, y^2) = 1 +2m^2(x^2 +y^2) +
 m^4(x^4 + y^4 - 2x^2 y^2)~, \label{defF}
 \ee
it is straightforward to show that
 \be
 {\cal D}z^{a'} {\cal D}z_{a'} - {\cal D}t^{\hat a}{\cal D}t^{\hat a} =
 F^{-2}(x^2,y^2)
 \left(dx^2 - dy^2\right) = F^{-2}\; dx^m dx_m~.
 \label{flat22}
 \ee
This relation demonstrates in which sense $ {AdS}_2\times S^2$ is conformal to
the Minkowski space $M_4$.

Another way to derive \p{flat22} (which directly applies to the supersymmetry
case) does not require passing to the coordinates $\tilde{x}^m = (z^{a'},
t^{\hat a})\,.$ It is as follows. One
 first examines how the differentials $dx^m$ are transformed under \p{tranprod}
 \bea
 && \delta_1 dx^{a'} = 2m^2 (a\cdot x) dx^{a'} -2m^2 (a^{a'} x^{b'} -
 x^{a'} a^{b'} ) dx_{b'} + 2 m^2 a^{a'} y^{\hat a} dy^{\hat a}~, \nn
 && \delta_1 dy^{\hat a} = 2m^2 (a\cdot x) dy^{\hat a} + 2m^2 y^{\hat a} x^{a'} dx_{a'}~, \nn
 && \delta_2 dy^{\hat a} = 2m^2 (c\cdot y) dy^{\hat a} -2m^2 (c^{\hat a} y^{\hat b} -
 y^{\hat a} c^{\hat b} ) dy^{\hat b} + 2 m^2 c^{\hat a} x^{a'} dx_{a'}~, \nn
 && \delta_2 dx^{a'} = 2m^2 (c\cdot y) dx^{a'} + 2m^2 x^{a'} c^{\hat a} dy^{\hat a}~.
 \label{12d}
 \eea
 One observes that, similar to other examples, $dx^m$ undergoes an
 $x^m$-dependent $SO(1,3)$ transformation and an
 $x^m$-dependent rescaling
 \be
 \delta d x^m = 2m^2\, [(a\cdot x) + (c\cdot y)] dx^m + L^{[mn]}_{(ind)} dx_n~.
 \ee
 Then one constructs a `semi-covariant' differential ${\cal D}x^m$ which
 undergoes only the induced Lorentz rotation
 \be
 {\cal D}x^m = f(x^2, y^2) dx^m~, \quad \delta f = - 2m^2[(a\cdot x) + (c\cdot
 y)]f~.
 \ee
 One gets a simple differential equation for $f$, which is solved, up to
 an overall integration constant, by
 \be
 f = F^{-1}\equiv \left[ 1 +2m^2(x^2 +y^2) +
                  m^4(x^4 + y^4 - 2x^2 y^2)\right]^{-{1\over 2}}\; ,
 \ee
 where $F$ is the quantity defined in \p{defF}. Thus we once again come to the
 conformal flatness relation \p{flat22}.

In conclusion of this Subsection, let us notice that all above formulas are
equally valid for an arbitrary $D$-dimensional Minkowski space $M_D\,.$ They
establish the conformal flatness, in the above sense, of the product spaces $
{AdS}_m \times  {S}^n = (z^{a'}, t^{\hat a})\,, \; m+n = D\,,$ with ${a'} = 0,1,...
m-1~, \;\; \hat a= m, ... D-1\,.$ The property that $ {AdS}_D$ is conformal to
$M_D$ is a corollary of this general statement. The necessary condition of the
conformal flatness is the equality of the $ {AdS}_m$ and $ {S}^n$ radii. It is
automatically satisfied when the $ {AdS}_m \times  {S}^n$ isometry group is
embedded into the conformal group of $M_D\,.$

\subsection{Supersymmetrization}
 Let us now pass to the supersymmetric case. We shall consider only
the four-dimensional case. First of all, we should identify the $D=4$
superconformal group, which might contain as a subgroup the isometry supergroup
$SU(1,1|2)$ or $OSp(4^*|2)$ of the $AdS_2\times S^2$ superspaces considered in
Subsection 2.2. It cannot be the $N=1$, $D=4$ superconformal group $SU(2,2|1)$,
since $SU(1,1|2)$, $OSp(4^*|2)$ and $SU(2,2|1)$ have the same number (eight) of
spinor generators, and they are obviously different.

The simplest $D=4$ superconformal algebra which may contain $su(1,1|2)$ or
$osp(4^*|2)$ is the $N=2$ superconformal algebra $su(2,2|2)$, its non-vanishing
(anti)commutation relations relevant to our study being
 \bea
 &&\{Q_\alpha^i, \bar Q_{\dot\alpha k}\} =
 2 \delta^i_k\sigma^m_{\alpha\dot\alpha} P_m~, \quad \{S_{\alpha k}, \bar S_{\dot\alpha}^i\} =
 2\delta^i_k \sigma^m_{\alpha\dot\alpha} K_m~, \nn
 && \{Q_\alpha^i, S^{\beta k}\} = \epsilon^{ik}
 (\sigma^{mn})_\alpha^\beta M_{mn} + 2i\epsilon^{ik}\delta^\beta_\alpha (D + iR) -4i
 \delta^\beta_\alpha T^{(ik)}~, \quad \mbox{and c.c.}~, \label{anticSC} \\
 && [K_m, Q^i_\alpha] = (\sigma_m)_{\alpha\dot\alpha} \bar S^{\dot\alpha i}~,
 \quad [P_m, S_{\alpha i}] = (\sigma_m)_{\alpha\dot\alpha} \bar Q^{\dot\alpha}_{i}~,
 \label{PS} \\
 &&[M_{mn}, P_s ] = i(\eta_{ns}P_m -\eta_{ms}P_n)~, \; [M_{mn}, Q^i_\alpha]
 = -{1\over 2}(\sigma_{mn})_\alpha^\beta Q^i_\beta~, \label{MP} \\
 && [T^{ij}, T^{lk}] = -i(\epsilon^{il} T^{jk} +  \epsilon^{jk} T^{il})~. \label{su2SC}
 \eea
The conjugation rules are as follows: $\overline{Q^i_\alpha} = \bar
Q_{\dot\alpha i}~,$ $\overline{S_{\alpha i}} = \bar S_{\dot\alpha}^i$.

Now we shall show that $osp(4^*|2)$ is a subalgebra of $su(2,2|2)$, while no an
appropriate subalgebra $su(1,1|2)$ can be found. It is straightforward to check
that the anticommutators of the following generators
 \bea
 && \hat{Q}^i_{1} = Q^i_1 + m\, S^i_1~, \quad \bar{\hat{Q}}_{\dot 1 i} =
 \bar Q_{\dot 1 i} - m\, \bar S_{\dot 1 i}~, \nn
 && \hat{Q}^i_{2} = Q^i_2 - m\, S^i_2~, \quad \bar{\hat{Q}}_{\dot 2 i}
 = \bar Q_{\dot 2 i} + m\, \bar S_{\dot 2 i} \label{22subalg}
 \eea
produce just the generators  \p{so12so3gen}
 \bea\label{su2}
 &&\{\hat Q^i_1, \bar{\hat Q}_{\dot 1 k}\} = 2\delta^i_k\left[ (P_0 +P_3) + m^2\,(K_0
 +K_3)
 \right]~,\phantom{AAAAAAAAAAAAAAAAAAAAAAAAA}\nn
 && \{\hat{Q}^i_2, \bar{\hat Q}_{\dot 2 k}\} = 2\delta^i_k \left[ (P_0 -P_3) +
 m^2\,(K_0-K_3)\right]~,\phantom{AAAAAAAAAAAAAAAAAAAAAAAAA} \nn
 && \{\hat{Q}^i_1, \bar{\hat Q}_{\dot 2 k}\} = 2 \delta^i_k\left[(P_1 - iP_2) - m^2\,
 (K_1 - iK_2)
 \right]~, \; \mbox{and c.c.}~,\phantom{AAAAAAAAAAAAAAAAAA} \nn
 && \{\hat{Q}^i_1, \hat Q^k_1 \} =\{\hat{Q}^i_2, \hat Q^k_2 \} = 0, \;
 \{\hat{Q}^i_1, \hat Q^k_2 \} = -4im \left[ \epsilon^{ik}\left(M_{03} + iM_{12}
 \right) + 2
 T^{(ik)} \right], \;\mbox{and c.c.}~.
 \eea
For establishing relation with the description of the $AdS_2\times S^2$ cosets
in Subsection 2.2, it is instructive to rewrite (\ref{su2}) in a  $D=4$ Lorentz
`covariant' fashion
 \bea\label{su22}
 &&\{\hat Q^i_\alpha, \bar{\hat Q}_{\dot \alpha k}\} = 2\delta^i_k\left
 [\sigma^{{a'}}_{\alpha\dot\alpha}
 (P_{a'}+m^2K_{a'})+\sigma^{\hat a}_{\alpha\dot\alpha}
 (P_{\hat a}-m^2K_{\hat a})\right]\;~,\nn
 && \{\hat{Q}^i_\alpha, \hat Q^k_\beta \}  =
 -4im \left[ \epsilon_{\alpha\beta}\epsilon^{ik}\left(M_{03} + iM_{12}
 \right) + 2(1-i)f_{\alpha\beta} T^{(ik)} \right], \;\mbox{and c.c.}\,,
 \eea
where ${a'}=0,3$ and $\hat a=1,2$ are the $AdS_2$ and $S^2$ vector
indices, respectively.

The commutators of the spinor charges with the  generators of $SO(2,1)$ and
$SO(3)$ are
 \bea\label{+}
& [P_{a'} + m^2\,K_{a'},\hat{Q}^i_\alpha ] =
 m\, \epsilon_{a'b'}\sigma^{b'}_{\alpha\dot\alpha}\bar{\hat Q}^{i\dot\alpha}~,
 \qquad
  [P_{\hat a} - m^2\,K_{\hat a},\hat{Q}^i_\alpha ] =
mi\, \epsilon_{\hat a\hat b}\sigma^{\hat b}_{\alpha\dot\alpha}\bar{\hat
Q}^{i\dot\alpha}\,,\nn
 &[M_{03},\hat{Q}^i_\alpha]=\sigma^{3~\beta}_{\alpha}\hat{Q}^i_\beta, \qquad
[M_{12},\hat{Q}^i_\alpha]=-i\sigma^{3~\beta}_{\alpha}\hat{Q}^i_{\beta}, \quad
{\rm and ~c.c.}~.
 \eea

Note that in addition to the generators of $so(2,1)$ and $so(3)$ the right hand
side of (\ref{su22}) also contains $su(2)$ generators $T^{(ik)}$.  The
subalgebra of $su(2,2|2)$ defined by (\ref{22subalg}) and (\ref{su22}) is a
superalgebra $osp\,(4^*|2)\,.$  The bosonic sector of this subalgebra is
$so^*(4)\oplus usp(2)\sim so(1,2) \oplus so(3) \oplus su(2)\,.$ The group
corresponding to the latter $su(2)$ acts only on the fermionic coordinates of
the relevant superspace (as in the $N=2,$ $D=4$ Minkowski superspace), it
commutes with special conformal transformations and, hence, does not affect the
geometry of the bosonic manifold which is always $ {AdS}_2\times S^2\,.$ We thus
conclude that the $ {AdS}_2\times S^2$ superspace can be realized as a
supercoset ${{OSp(4^*|2)}\over{SO(1,1)\times SO(2)\times SU(2)}}\,.$ The
$su(1,1|2)$ superalgebra is not a subalgebra of $osp\,(4^*|2)$ and it cannot be
obtained from the latter by any contraction. We conclude that no $su(1,1|2)$
subalgebra exists in $su(2,2|2)\,$, such that its bosonic subalgebra
$so(2,1)\oplus so(3)$ lies in the bosonic conformal subalgebra $so(2,4)$ of
$su(2,2|2)\,$.\footnote{Of course there is an obvious embedding of $su(1,1|2)$ in
$su(2,2|2)$ such that their internal $su(2)$ sectors coincide. However, this
embedding does not suit our purposes.}

Let us now present the basic anticommutation relations \p{su2} in a form where
the $so(2,1)\oplus so(3)\oplus su(2)$ structure is manifest. We introduce
$$
Q^{A ii'}~, \quad \overline{(Q^{A ii'})} = \epsilon_{ik}\epsilon_{i'k'}Q^{A
kk'}~,
$$
such that \be Q^{A=1, i,i'=1} \equiv -i\hat Q_2^i~, \quad Q^{A=2, i,i'=2} =
-\hat Q^i_1~, \quad Q^{1i2} = i\bar{\hat Q}^i_{\dot 2},\quad Q^{2i1}=\bar{\hat
Q}^i_{\dot 1}\,,
 \ee
 where $A, i$ and $i'$ are spinor
indices of $SL(2,R)\sim SO(2,1)$ and of two $SU(2)$ groups, respectively. Then
\p{su2} can be rewritten in the following concise form
 \be \{ Q^{A ii'}, Q^{Bkk'} \} = -4m(\,\epsilon^{AB}\epsilon^{ik}T^{i'k'}_1 +
\epsilon^{ik}\epsilon^{i'k'}T^{AB}_2 -2 \epsilon^{AB}\epsilon^{i'k'}T^{ik}\,)~,
\label{symmform}
 \ee
 where all bosonic generators satisfy the
same commutation relations \p{su2SC}, $T_2^{AB}$ and $T_1^{i'k'}$ being
generators of $SO(2,1)$ and $SO(3)$, respectively. They are related to the
original ones as
 \bea && T_1^{11} = {i\over 2m}[(P_1 + iP_2) -m^2(K_1 +iK_2)]~,
\; T_1^{22} = (T_1^{11})^\dagger~, \; T_1^{12} = -iM_{12}~, \nn && T_2^{11} =
-{1\over 2m}[(P_0 - P_3) +m^2(K_0 - K_3)]~, \; T_2^{22} = -{1\over 2m}[(P_0 +
P_3) +m^2(K_0 +K_3)]~, \nn && T_2^{12} = M_{03}~.
 \eea
 In this notation it is easy
to see that $osp(4^*|2)$ is a particular case of a real form of the exceptional
Lie superalgebra $D(2,1;\alpha)\,$. The basic anticommutation relation of the
latter \cite{sorba} can be obtained by replacing the coefficients before the
generators $T_1$ and $T$ in the r.h.s. of \p{symmform} by numerical parameters
$\alpha$ and $-(1+\alpha)\,$, respectively,
 \be\label{d(2,1)}
\{ Q^{A ii'}, Q^{B kk'} \} = -4m(\,\alpha\,\epsilon^{AB}\epsilon^{ik}T^{i'k'}_1
+ \epsilon^{ik}\epsilon^{i'k'}T^{AB}_2 -(1+\alpha)\,
\epsilon^{AB}\epsilon^{i'k'}T^{ik}\,)~.
 \ee
The superalgebra $osp(4^*|2)$ is recovered from (\ref{d(2,1)}) with the choice
$\alpha =1\,,$ while the choices $\alpha = 0$ or $\alpha = -1$ lead to two
isomorphic superlagebras, each being a semidirect sum of a superalgebra
$su(1,1|2)$ and an external automorphism algebra $su(2)$. The $su(1,1|2)$
superalgebra corresponding to $\alpha=-1$ is obtained from (\ref{su22}) by
skipping the $SU(2)$ generators $T^{ij}$ and changing the sign in front of the
$SO(3)$ generators $P_{\hat a}-m^2K_{\hat a}$ and $M_{12}\,,$ the commutation relations
(\ref{+}) being unchanged. This form of the $su(1,1|2)$ superalgebra
corresponds to the form of the Maurer--Cartan equations analyzed in Subsection
2.2.

 To learn how the conformal flatness relation \p{flat22}
generalizes to the supersymmetry case, we need, before all, to have the
realization of the subalgebra \p{su2} on the coordinates of the $N=2, ~D=4$
superspace. We shall proceed from the $N=2,~D=4$ superconformal transformations
\p{N2eps}, \p{N2eta}, \p{N2scint} in the left-chiral parametrization. It will
be convenient to relabel the coordinates and $N=2$ supersymmetry parameters as
follows
 \bea
 && x^{1\dot 1}_L \equiv x^{++}~,\; x^{2\dot 2}_L \equiv x^{--}~, \;x^{2\dot 1}_L
 \equiv
 z~,\;x^{1\dot 2}_L \equiv \bar z~, \nn
 && \theta^{1}_i \equiv \theta^{+}_i~, \; \theta^{2}_i \equiv \theta^{-}_i~, \;
 \bar\theta^{\dot 1i} = \bar\theta^{+ i}~, \;\bar\theta^{\dot 2i} \equiv
 \bar\theta^{- i}~, \nn
 && \epsilon^{1}_i \equiv \epsilon^{+}_i~, \; \epsilon^{2}_i \equiv \epsilon^{-}_i~,
 \;
 \bar\epsilon^{\dot 1i} = \bar\epsilon^{+ i}~, \;\bar\epsilon^{\dot 2i} \equiv
 \bar\epsilon^{- i}~.
 \eea
The subalgebra \p{22subalg} is singled out by the following identification of
the parameters of the special superconformal and Poincar\'e supersymmetry
transformations (this choice properly breaks $D=4$ Lorentz invariance):
 \be
 \eta^1_i = - m \epsilon^+_i~, \;
 \eta^2_i = m \epsilon^-_i~,\;
 \bar\eta^{\dot 1i} = m \bar\epsilon^{+i}~, \;
 \bar\eta^{\dot 2i} = - m \bar\epsilon^{- i}~.
 \ee
The corresponding transformations are given by
 \bea
 && \delta x^{++} = -4i \bar \epsilon^{+}\theta^+ -4i m \left[x^{++} \, (\epsilon^-
 \theta^+)
 + z\,(\epsilon^+\theta^+) \right]~, \nn
 && \delta x^{--} = -4i \bar \epsilon^{-}\theta^- -4i m \left[x^{--} \, (\epsilon^+
 \theta^-)
 + \bar z\,(\epsilon^-\theta^-) \right]~, \nn
 && \delta z  = -4i \bar \epsilon^{+}\theta^- -4i m \left[x^{++} \, (\epsilon^-
 \theta^-) + z\,(\epsilon^+\theta^-) \right]~, \nn
 && \delta \bar z  = -4i \bar \epsilon^{-}\theta^+ -4i m \left[x^{--} \,
 (\epsilon^+
 \theta^+) + \bar z\,(\epsilon^-\theta^+) \right]~, \nn
 && \delta \theta^+_i = \epsilon^+_i + 2im \left[  \epsilon^-_i
 \,(\theta^+)^2 - 2 (\epsilon^+\theta^+)\theta^-_i \right] - m\left( \bar\epsilon^-_i
 x^{++} +
 \bar\epsilon^+_i \bar z \right)~, \nn
 && \delta \theta^-_i = \epsilon^-_i + 2im \left[  \epsilon^+_i
 \,(\theta^-)^2 - 2 (\epsilon^-\theta^-)\theta^+_i \right] - m\left( \bar\epsilon^+_i
 x^{--} +
 \bar\epsilon^-_i z \right) \label{22susytran}
 \eea
 (for brevity, we
have omitted the chiral index `L' on the coordinates and denoted
$ a^ib_i\equiv (ab)$). Note an asymmetry in the transformation laws of
 $z$ and $\bar z$ which is of course related to the fact that these
 coordinates are not mutually conjugate in the complex chiral basis.
 The transformation of the super Poincar\'e covariant differential
 $\Pi^{\alpha \dot\alpha}$ under this subgroup, up to an induced Lorentz
 rotation, is as follows
 \be
 \delta \Pi^{\alpha\dot\alpha} = - \left[ 2im \left(\epsilon^-\theta^+
 + \epsilon^+\theta^- \right) + \mbox{c.c.} \right] \Pi^{\alpha\dot\alpha}~.
 \label{22weightsus}
 \ee

 As in the bosonic case, in order to construct the `semi-covariant'
 differential which would undergo only coordinate dependent Lorentz
 rotations under the action of the $ {AdS}_2 \times  {S}^2$ supergroup, one
 should
 construct a density $B(x,\theta, \bar\theta)$ which  compensates the
 weight factor in \p{22weightsus},
 \be
 \delta B =\left[ 2im \left(\epsilon^-\theta^+
 + \epsilon^+\theta^- \right) + \mbox{c.c.} \right] B~.
 \ee
 In analogy to the previous supersymmetric examples, we assume $B$ to have the
 product structure
 \be
 B(x, \theta, \bar\theta) = B_L(x_L, \theta)B_R(x_R, \bar\theta)~, \;
 \delta B_L =2im \left( \epsilon^-\theta^+
 + \epsilon^+\theta^- \right)\,B_L~, \;\; \delta B_R = \overline{\delta
 B_L}~. \label{susweight22}
 \ee
 The most general $\theta $ expansion of $B_L$ compatible with the
 $AdS_2\times S^2$ isotropy group $SO(1,1)\times SO(2)$ is as follows
 \bea
 B_L &=& a(X, Y) + b(X,Y)\theta^+\theta^- + d(X,Y)x^{--}z(\theta^+)^2
 + c(X,Y)x^{++}\bar z(\theta^-)^2 \nn
 && +\, f(X,Y) (\theta^+)^2  (\theta^-)^2~,
 \label{expan22} \\
 X &\equiv & x^{++}x^{--}~, \quad Y = z\bar z~.\label{XY}
 \eea

 Requiring $B_L$ to have the transformation rule \p{susweight22} under
 \p{22susytran}, we find quite a lot of equations for the coefficient
 functions in \p{expan22}. But only few of them are essential
 \bea
 4i \frac{\partial a}{\partial X} + m b + 2m Y d &=& 0~, \nn
 4i \frac{\partial a}{\partial Y} + m b + 2m X d &=& 0~, \nn
 4im \left( X \frac{\partial a}{\partial X} + Y\frac{\partial a}{\partial Y} \right)
 - b + 2i m a &=& 0~, \nn
 4im \left(\frac{\partial a}{\partial X} + \frac{\partial a}{\partial Y} \right)
 - 2 d &=& 0~, \nn
 i \frac{\partial b}{\partial X} - 2i Y\frac{\partial d}{\partial Y}
 - 2i d - m f &=& 0~, \nn
 c &=& d~, \label{eqs22}
 \eea
 while the rest of equations become identities on the solutions of these
 basic ones (and so serve as self-consistency conditions). It is straightforward
 to find that (up to an arbitrary overall
 constant) the general solution of \p{eqs22} is given by
 \bea
 && a(X,Y) = \frac{1}{[1 + 2m^2(X+Y) + m^4(X-Y)^2]^{1/4}}~, \nn
 && b(X,Y) = 2im \frac{1 + m^2(X+Y)}{[1 + 2m^2(X+Y) + m^4(X-Y)^2]^{5/4}}~, \nn
 && c(X,Y) = d(X,Y) = -2im^3\,\frac{1}{[1 + 2m^2(X+Y) + m^4(X-Y)^2]^{5/4}}~, \nn
 && f(X,Y) = - \frac{m^2}{[1 + 2m^2(X+Y) + m^4(X-Y)^2]^{5/4}}~. \label{solv22}
 \eea
 Thus
 \be
 {\cal D}x^{\alpha\dot\alpha} = B_L B_R \Pi^{\alpha\dot\alpha}
 \ee
 and the supersymmetric analog of the bosonic conformal flatness relation \p{flat22}
 is
 \be
 {\cal D}s^2 = {\cal D}x^{\alpha\dot\alpha}{\cal D}x_{\alpha\dot\alpha} =
 (B_L)^2(B_R)^2
 (\Pi^{\alpha\dot\alpha} \Pi_{\alpha\dot\alpha})~. \
 \ee

We have thus found the explicit superconformal factors in the expressions for
the supervielbeins of the $AdS_2\times S^2$
conformally flat superspace ${{OSp(4^*|2)}\over{SO(1,1)\times SO(2)\times
SU(2)}}$ discussed in Section 2. These factors are defined by
eqs. \p{expan22}, \p{XY}, \p{solv22} and their conjugate.

\setcounter{equation}{0}

\section{Applications -- an outlook}
We shall now demonstrate, using the results obtained, that in the
superconformally flat superbackgrounds the classical dynamics of a massless
particle is (conformally) equivalent to the classical dynamics of a
superparticle in flat superspace.\footnote{In \cite{bls} the equivalence of the
dynamics of a massless bosonic particle propagating in conformally flat
backgrounds has been demonstrated in a twistor--like formulation. Using the
results of this Section one can show that the equivalence established in
\cite{bls} can be extended to massless $AdS$ $D=4$ superparticles formulated in
the supertwistor framework.}

The classical action of a massless superparticle in a $D$--dimensional
supergravity background parametrized by supercoordinates
$z^M=(x^m,\theta^\mu,\bar\theta^{\dot\mu})$ and described by supervielbeins
\linebreak
 $dz^M\,E^A_M(x,\theta,\bar\theta)$, $A=(a,\alpha,\dot\alpha)$, has
the following form
\begin{equation}\label{5.1}
S={1\over 2}\int\, d\tau\,{1\over{e(\tau)}}\dot z^M\,\dot z^N\,E^a_M \,E^b_N
\,\eta_{ab}~,
\end{equation}
where $\dot z^M={d{z^M}\over{d\tau}}$ and $e(\tau)$ is a Lagrange multiplier
which insures the mass shell condition $\dot z^M\,E^a_M \, \dot
z^N\,E^b_N\eta_{ab} =0~.$

The action (\ref{5.1}) is invariant under worldline reparametrizations and
under a local $\kappa$-symmetry, provided the supervielbeins satisfy
appropriate supergravity constraints (for a review of superparticle models see
\cite{Sorokin:1999jx}). In particular, the superbackground can be of the
$AdS\times S$ types discussed in Sections 2, 3 and 4. Then the supervielbeins
$dz^M E^a_M$ are as in equations (\ref{superflat}), (\ref{cf}), (\ref{Pi}) and
(\ref{2cfEa}), and the action (\ref{5.1}) takes the form
\begin{equation}\label{5.2}
S={1\over 2} \int\,
d\tau\,{{e^{4\Phi(z)}}\over{e(\tau)}}\Pi^a\Pi^b\,\eta_{ab}\,.
\end{equation}
Note that the action (\ref{5.2}) is invariant with respect to the
superconformal transformations of the target superspace coordinates, provided
the Lagrange multiplier  $e(\tau)$ (einbein) gets rescaled in an appropriate
way. The property of the actions of massless bosonic particles, spinning
particles and superparticles to be target--space (super)conformal invariant is
actually well known and has been extensively discussed in the literature (see
e.g. \cite{mar}--\cite{bars1}).  Our observation is that if the superbackground
is superconformally flat, the conformal factor can be absorbed into the
redefined einbein
\begin{equation}\label{e'}
e'(\tau)=e^{-4\Phi(z)}e(\tau)\,,
\end{equation}
and the action reduces to that of a massless particle in a flat
superbackground. We thus conclude that, for instance, the dynamics of massless
superparticles propagating on the $AdS_4$ coset superspace
${{OSp(2|4;R)}\over{SO(1,3)\times SO(2)}}\,,$ on the $AdS_2\times S^2$ coset
superspace ${{OSp(4^*|2)}\over{SO(1,1)\times SO(2)\times SU(2)}}$ and in flat
$N=2$, $D=4$ superspace are classically equivalent (at least locally), the
superconformal group being $SU(2,2|2)$ for all three cases. This essentially
simplifies the analysis of the superparticle mechanics.

At the quantum level,
 because of the operator ordering ambiguity problem
and of the nontrivial topological structure of the $AdS\times S$ manifolds,
these cases should show up differences from the flat one. For instance, the
operator ordering for each of these cases should be fixed (at least partially)
by the requirement that the quantum constraints ($\sim$ superfield equations)
are covariant with respect to the superconformal transformations and, moreover,
possess a symmetry associated with the isometry of the specific $AdS$
superbackground.\footnote{This requirement is well known both in the
conventional quantum field theory (see e.g. \cite{bogo})  and quantum string
theory \cite{gsw}. It is usually applied to justify the consistency of the
quantum theory by constructing the generators of all the symmetries in terms of
quantum variables (fields) and verifying that they still satisfy the
commutation relations characteristic of the symmetries of the classical
theory.} Such a requirement will obviously result in different equations of
motion for superfields which describe the first quantized state vectors.  In
addition, the definition of energy and of the mass of states on
$AdS$--manifolds is subtle (see Duff {\it et al} \cite{kk} for a review).

As an example let us consider in more details the operator ordering procedure
for the canonical quantization of a bosonic massless particle in flat and $AdS$
spaces.

 From the bosonic counterpart of the action (\ref{5.2}) it follows that,
classically, in both cases we have a single first--class constraint
$p_mp^m=0\,,$ $p_m$ being the canonical momentum of the particle. The classical
equations of motion of the massless particle in flat space and in $AdS$ can be
made equivalent by rescaling the einbein $e(\tau)\,$. And in this sense the
dynamics in flat space and in $AdS$ space are equivalent.  What is different is
the symmetry of the dynamical systems in flat and $AdS$ spaces, because of
different geometrical properties of these backgrounds.

When we quantize these systems we must respect these symmetries, and this is a
criterion for the choice of operator ordering. Let us consider how it works. We
start with the constraint $p_mp^m=0\,$, which upon quantization we would like to
apply on the wave function and to obtain the field equation of motion.

If we are in flat space, we directly convert $p_mp^m$ into the Klein-Gordon
operator by replacing $p_m$ by $-i\partial_m\,$. In this way we get the correct
Klein-Gordon equation in flat space which respects the Poincar\'e symmetry of the
initial classical system.

However, if we are in an $AdS$ background, we cannot proceed in the same way,
since $\partial_m\partial^m$ is not invariant under isometries of the $AdS$
space. To obtain the correct $AdS$--invariant Klein--Gordon equation we should
use a different operator ordering procedure based on the insertion of the
conformal factor $exp{\,(2(D-2)\phi(x))}$. To understand how it works let us
start with the result. Upon quantization we should get the Klein-Gordon
equation in the form
$$
D_m g^{mn}\partial_n V(x)=0~,
$$
where $D_m$ is the $AdS$ covariant derivative and $g^{mn}$ is the inverse $AdS$
metric. The above equation can be rewritten in the equivalent form as
\begin{equation}\label{5.4}
g^{-{1\over 2}}\partial_m \left(g^{1\over 2}g^{mn}\partial_n V(x)\right)=0~,
\quad g\equiv\det g_{mn}\,.
\end{equation}
Now let $g_{mn}=e^{4\phi(x)}\eta_{mn}~.$ Then the equation (\ref{5.4}) reduces
to
\begin{equation}\label{5.5}
 e^{-2D\phi}\partial_m \left(e^{2(D-2)\phi} \partial^m V(x)\right)=0~,
\end{equation}
where $D$ is the dimension of the $AdS$ background.

 The form of these equations suggests which kind of the operator ordering
procedure we should follow when quantizing the $AdS$ particle. We should
rewrite $p_mp^m$
 in the following classically equivalent form
\begin{equation}\label{5.6}
\eta^{mn}p_mp_n\equiv p_mp^m= e^{-2(D-2)\phi}p_me^{2(D-2)\phi}p^m\,.
\end{equation}
Upon quantization we shall impose the first class constraint (\ref{5.6}) on the
state vector $V(x)$ in the coordinate representation ($x^m\rightarrow x^m\;$,
$\; p_m\rightarrow -i \partial_m$). Taking the operator ordering of the
constraint as in the right hand side of (\ref{5.6}), we get the $AdS$--covariant
field equation, which is obviously different from the flat space Klein-Gordon
equation. This equation differs from (\ref{5.5}) by the factor $e^{-4\phi}$
which can be restored already at the classical level from the requirement of
the invariance (versus covariance) of the constraints with respect to the
isometries of the $AdS$ space, $p_mp^m =0 \; \rightarrow e^{-4\phi} p_mp^m=0\,.$

By making a different momentum ordering in (\ref{5.6}) such that for $D > 2$
\begin{eqnarray}\label{rorder}
p_mp^m&=&e^{-2(D-2)\phi}p_me^{2(D-2)\phi}p^m \nonumber\\
&&  - 4c\;{{D-1}\over D-2}\left[e^{-2(D-2)\phi}
p_me^{2(D-2)\phi}p^m-e^{-(D-2)\phi}p_mp^me^{(D-2)\phi}
\right]
\end{eqnarray}
and for $D=2$
\begin{eqnarray}\label{rorder2}
p_mp^m&=&p_mp^m  - c\;\left[e^{-\phi} p_me^{\phi}p^m-p_me^{-\phi}p^me^{\phi}
\right]\,,
\end{eqnarray}
one gets a contribution to the Klein--Gordon equation (\ref{5.4}) proportional
to the scalar curvature ${\cal R}$ of the conformally flat manifold
\begin{equation}\label{r}
(D_m g^{mn}\partial_n -  c{\cal R}) V(x)=0\,.
\end{equation}
The term in the square brackets of eqs. (\ref{rorder}) and (\ref{rorder2}),
which classically equals to zero, produces the scalar curvature term ${\cal R}$
when acting on $V(x)$ upon quantization. In the case of the AdS spaces with
constant ${\cal R}$ such a term modifies the mass operator of the fields. The
arbitrary constant $c$ can be fixed by requiring eq. (\ref{r}) to be
conformally invariant
\cite{fronsdal}, for instance $c= {1\over 6}$ in $D=4$ .

Hence the quantum dynamics of the massless particle in flat space and in $AdS$
space are not equivalent.

The same reasoning  applies to the supersymmetric case. Note that at the
classical level not only the mass shell condition $p_mp^m=0\,,$ but also the
fermionic constraints $\pi_\alpha-ip_m\gamma^m_{\alpha\beta}\theta^\beta=0$
(where $\pi_\alpha$ is the momentum conjugate to $\theta^\alpha$) are
equivalent for the massless superparticles (\ref{5.2}) moving in  different
superconformally flat backgrounds.

Quantum equivalence of the dynamics of massless (super)particles in different
conformally flat (super)backgrounds can be also analyzed in the framework of
path integral quantization. In a simpler model of non--relativistic bosonic
particles moving in a curved Riemann space with a metric $g_{mn}$ one performs
quantization by taking the path integral over the trajectories $x_m(t)$ with a
functional measure $(\det{g_{mn})^{1\over 2}}$ invariant under target space
diffeomorphisms (see e.g. \cite{abers,bastia}). Such a covariant measure in the
coordinate path integral can be derived from the path integral over phase space
trajectories by integrating over the momentum paths. For the relativistic
superparticle (\ref{5.1}) moving in a curved superbackground described by the
supervierbein $E_M^A\,$, because of the presence of constraints, the quantum
theory will be defined by a generalized  path integral over a
Batalin--Fradkin--Vilkovisky--extended phase space.  The functional integral
over superspace trajectories $z_M({\tau})$ invariant under the target space
superdiffeomorphisms should emerge upon doing an integral over the
BFV--extended momentum trajectories, which should provide the measure with a
covariant factor $\mbox{Ber}(E_M^A)\,$. The derivation of this measure is an
interesting and still unsolved problem. The above reasoning suggests that the
functional integration measure of the quantum supersymmetric theory based on
the classical action (\ref{5.1}) includes $\mbox{Ber}(E_M^A)\,$. At the next step,
respecting the symmetries of the classical model in each of the
superconformally flat backgrounds, one should choose a suitable  regularization
prescription, e.g. a suitable discretization of superpaths, and operator ordering
in the definition of this formal functional integral (for non--supersymmetric
case see \cite{parker,kuchar}).

In the case of massive superparticles, superstrings and superbranes the action
does not have invariances corresponding to {\it superconformal} symmetries of
target superspace \cite{abra}. It depends not only on the vector components of
the target space supervielbeins but also on the superform gauge fields ($C_1$
for massive superparticles, $B_2$ for superstrings, $B_2$, $C_0$, $C_2$, $C_4$
for super--D$p$--branes with $p=1,3$, etc.), whose pull--back enters the
Wess--Zumino term. Their field strengths are expressed through bosonic and
fermionic supervielbeins (\ref{superflat}) as a consequence of relevant target
space supergravity constraints. It would be of interest to examine whether the
superconformally flat structure of the target space supervielbeins may result
in a simplification of actions for such objects propagating in $AdS$
supermanifolds.

Superconformal quantum mechanics (of multi--black holes) with a $D(2,1;\alpha)$
(and, in particular, $OSp(4^*|2)$) superconformal symmetry has been considered
in \cite{strominger} as a generalization of the $SU(1,1|2)$ superconformal
mechanics in the background of a single Reissner--Nordstr\"om black hole. We
therefore see that the supergroups $D(2,1;\alpha)$ have appeared in physical
applications. Hence, it would be of interest to study whether they are relevant
to supergravity theories and, in particular, whether coset superspaces
${{D(2,1;\alpha)}\over{SO(1,1)\times SO(2)\times SU(2)}}$ and
${D(2,1;\alpha)\times D(2,1;\alpha)}\over{SO(1,2)\times SO(3)\times SU(2)\times
SU(2)}$ can be recovered as solutions of some $N$ -- extended  (conformal)
supergravities with local $SU(2)$ and $SU(2)\times SU(2)$ R--symmetries.

\section{Conclusion}
We have analyzed the superconformal structure of a class of supermanifolds with
the $AdS\times S$  bosonic body and proposed a recipe for deriving the exact
form of the conformal factors of the supervielbeins. In particular, we have
demonstrated that the $AdS_4$ and $AdS_2\times S^2$ superspaces whose
superconformal group is $SU(2,2|2)$ are superconformally flat and, hence,
conformally--equivalent to $N=2$, $D=4$ flat superspace. This is also the case
for the $AdS_3\times S^3$ superspace associated with the supercoset
${OSp(4^*|2)\times OSp(4^*|2)}\over{SO(1,2)\times SO(3)\times SU(2)\times
SU(2)}$ which is conformally equivalent to flat $N=(2,0)$, $D=6$ superspace.
However, the conventional $AdS_{D\over 2}\times S^{D\over 2}$ superspaces which
are maximally supersymmetric solutions of classical $N=2$, $D=4,6,10$
supergravity constraints, are not conformal supermanifolds, since for the
reasons explained in Sections 2 and 4, the isometry supergroups of these
supermanifolds are not subgroups of the superconformal groups in D=4,6 and 10
dimensions, respectively. Therefore, these $AdS\times S$ vacuum configurations
are not superconformal and the issue of their stability under higher order
corrections to the quantum effective action of the complete supersymmetric
theory should be revised (as e.g. in \cite{kallosh1}).

Let us note that the `pure' $N$--extended $AdS_2$  coset superspaces, i.e. the
ones without `$S$--factors' are always superconformal. This is because the
isometry supergroups of such supercosets can always be embedded into an
appropriate $N$--extended $D=2$ superconformal group, which is in agreement
with the conclusion of \cite{Howe} about the superconformally flat structure of
$D=2$ supergravities.

`Pure' $N$--extended $AdS_3$ coset superspaces are superconformal when their
isometries can be embedded into the $N$--extended $D=3$ superconformal group
$OSp(N|4;R)$. Examples are the $N=1$ supercoset ${OSp(1|2;R)\times
SO(1,2)}\over {SO(1,2)}\,$, $N=2$ supercosets ${OSp(1|2;R)\times
OSp(1|2;R)}\over {SO(1,2)}$ and \linebreak ${OSp(2|2;R)\times
SO(1,2)}\over{SO(1,2)\times SO(2)}\,$ $ \sim \,$ $ {{SU(1,1|1)\times
SL(2;R)}\over{SL(2;R)\times U(1)}}$, and, in general, the
supercosets ${OSp(N-n|2;R)\times
OSp(n|2;R)}\over {SO(1,2)\times SO(N-n)\times SO(n)}$.

The $AdS_5$ coset superspace ${SU(2,2|N)}\over {SO(1,4)\times U(N)}$ is
superconformal only for $N=1$, since only $SU(2,2|1)$ can be embedded as a
subgroup into the unique $D=5$ superconformal group $F(4;2)\,$,
with R--symmetry being $USp(2)\sim SU(2)$ \cite{fs}. For
$D \geq 5$ we have not found  $AdS_D$ superspaces, whose isometries could be
embedded into the corresponding superconformal group. For instance, the $AdS_6$
supercoset ${F(4;2)}\over{SO(1,5)\times USp(2)}$ is not superconformal, since
$F(4;2)$ is not a subgroup of the $D=6$ superconformal group $OSp(8^*|4)$
\cite{fs}.

One may conjecture that in higher dimensions, and in particular in $D=10$,
there exist generalized $AdS_m\times S^n$ superspaces enlarged with tensorial
charge coordinates and, possibly, with additional Grassmann--odd coordinates,
which can presumably be superconformally flat with respect to generalized
superconformal groups. Thus the study of superconformal structure of higher
dimensional superspaces brings us to supersymmetric models with tensorial
central charge coordinates (see e.g. \cite{45,rudichev,bars,46}) which
previously have already found various other motivations for their
consideration, including exotic BPS configurations which preserve more than one
half supersymmetry \cite{bl,bls,zima,BAIL} and the theory of higher spin fields
\cite{vasiliev}. This point still requires a detailed analysis. \vspace{0.3cm}

{\bf Acknowledgements}. D.S. is grateful to N. Berkovits, M. Tonin and P.
Pasti, and E.I. and J.L. thank I. Buchbinder and K. Stelle for interest in this
work and discussions. This work was partially supported by the European
Community's Human Potential Programme under contract HPRN-CT-2000-00131 Quantum
Spacetime (D.S.), by the Grant N 383 of the Ukrainian State Fund for
Fundamental Research (I.B. and D.S.) by the INTAS Research Project N 2000-254
(I.B., E.I. and D.S.), by the DGICYT research grant PB-96-0756 (I.B.) and by
the KBN grant 5P03B05620 (J.L.). E.I.  acknowledges partial support from the
Bogoliubov--Infeld program, the grants RFBR--DFG No 02-02-04002, DFG 436 -- RUS
113/669 and RFBR--CNRS No 01-02-22005.



\begin{thebibliography}{99}
\bibitem{kahler}
E.~E.~Donets, A.~Pashnev, V.~O.~Rivelles, D.~Sorokin and M.~Tsulaia,
Phys.\ Lett.\ B {\bf 484}, 337 (2000) [arXiv:hep-th/0004019];
Czech.\ J.\ Phys.\  {\bf 50}, 1215 (2000).
\bibitem{kk}
M.~J.~Duff, B.~E.~Nilsson and C.~N.~Pope,
Phys.\ Rept.\  {\bf 130}, 1 (1986).
\\
D.~P.~Sorokin and V.~I.~Tkach,
Sov. J. Part. Nucl. {\bf 18}, 441 (1987).
\bibitem{ads/cft}
J.~Maldacena, Adv. Theor. Math. Phys. {\bf 2}, 231 (1998).
[arXiv:hep-th/9711200].
\\
S.~S.~Gubser, I.~R.~Klebanov and A.~M.~Polyakov, Phys. Lett. {\bf B428}, 105
(1998). [arXiv:hep-th/9802109].
\\
E.~Witten, Adv. Theor. Math. Phys. {\bf 2}, 253 (1998). [arXiv:hep-th/9802150].
\\
O.~Aharony, S.~S.~Gubser, J.~Maldacena, H.~Ooguri and Y.~Oz,
Phys.\ Rept.\  {\bf 323}, 183 (2000) [arXiv:hep-th/9905111].
\bibitem{qm}
V.~P.~Akulov and A.~I.~Pashnev,
Theor.\ Math.\ Phys.\  {\bf 56} (1983) 862 [Teor.\ Mat.\ Fiz.\ {\bf 56} (1983)
344].
\\
S.~Fubini and E.~Rabinovici,
Nucl.\ Phys.\ B {\bf 245}, 17 (1984).
\\
E.~A.~Ivanov, S.~O.~Krivonos and V.~M.~Leviant,
J.\ Phys.\ A {\bf 22}, 4201 (1989).
\\
J.~A.~de Azcarraga, J.~M.~Izquierdo, J.~C.~Perez Bueno and P.~K.~Townsend,
Phys.\ Rev.\ D {\bf 59}, 084015 (1999) [arXiv:hep-th/9810230].
\bibitem{black}
P.~Claus, M.~Derix, R.~Kallosh, J.~Kumar, P.~K.~Townsend and A.~Van Proeyen,
Phys.\ Rev.\ Lett.\  {\bf 81}, 4553 (1998) [arXiv:hep-th/9804177].
\bibitem{cflat}
P.~Claus, R.~Kallosh, J.~Kumar, P.~K.~Townsend and A.~Van Proeyen,
JHEP {\bf 9806}, 004 (1998) [arXiv:hep-th/9801206].
\bibitem{abz}
V.~P.~Akulov, I.~A.~Bandos and V.~G.~Zima,
Theor.\ Math.\ Phys.\  {\bf 56} (1983) 635 [Teor.\ Mat.\ Fiz.\ {\bf 56} (1983)
3].
\bibitem{pesando}
I.~Pesando,
JHEP {\bf 9902}, 007 (1999) [arXiv:hep-th/9809145].
\bibitem{berkovits}
N.~Berkovits, M.~Bershadsky, T.~Hauer, S.~Zhukov and B.~Zwiebach,
Nucl.\ Phys.\ B {\bf 567}, 61 (2000) [arXiv:hep-th/9907200].
\bibitem{zhou}
M.~Kreuzer and J.~G.~Zhou,
Phys.\ Lett.\ B {\bf 472}, 309 (2000) [arXiv:hep-th/9910067].
\bibitem{bg}
T.~Banks and M.~B.~Green,
JHEP {\bf 9805}, 002 (1998) [arXiv:hep-th/9804170].
\bibitem{kallosh1}
R.~Kallosh, ``Black holes, branes and superconformal symmetry'',
hep-th/9901095.
\bibitem{osp164}
S.~Ferrara and M.~Porrati,
Phys.\ Lett.\ B {\bf 458}, 43 (1999) [arXiv:hep-th/9903241].
\\
R.~D'Auria, S.~Ferrara and M.~A.~Lled\'o,
Lett.\ Math.\ Phys.\  {\bf 57}, 123 (2001) [arXiv:hep-th/0102060].
\\
S.~Ferrara and M.~A.~Lled\'o, ``Considerations of Super Poincar\'e Algebras and
their Extensions to Simple Superalgebras'', hep-th/0112177.
\bibitem{fs}
S.~Ferrara and E.~Sokatchev, ``Conformally coupled supermultiplets in four and
five dimensions'',  hep-th/0203068.
\bibitem{bars}
I.~Bars, C.~Deliduman and D.~Minic,
Phys.\ Lett.\ B {\bf 457}, 275 (1999) [arXiv:hep-th/9904063].
\bibitem{west}
P.~West,
JHEP {\bf 0008}, 007 (2000) [arXiv:hep-th/0005270].
\bibitem{Howe}
P.~S.~Howe and R.~W.~Tucker,
Phys.\ Lett.\ B {\bf 80}, 138 (1978).
\\
P.~S.~Howe,
J.\ Phys.\ A {\bf 12}, 393 (1979).
\bibitem{sconf}
W.~Nahm, Nucl. Phys. B {\bf 135}, 149 (1978).
\\
Z.~Hasiewicz, J.~Lukierski, P.~Morawiec, Phys. Lett. B {\bf 130}, 55 (1983).
\bibitem{sorba}
L.~Frappat, P.~Sorba and A.~Sciarrino, ``Dictionary on Lie superalgebras'',
hep-th/9607161.
\bibitem{Claus:1998us}
P.~Claus, R.~Kallosh and A.~Van Proeyen, ``Conformal symmetry on world volumes
of branes'', hep-th/9812066.\\
A.~Van Proeyen, ``Tools for supersymmetry'', hep-th/9910030.
\bibitem{IIB}
P.~S.~Howe and P.~C.~West,
Nucl.\ Phys.\ B {\bf 238}, 181 (1984).
\bibitem{RK.98}
R.~Kallosh and A.~Rajaraman,
Phys.\ Rev.\ D {\bf 58}, 125003 (1998) [arXiv:hep-th/9805041].
\bibitem{ferber}
A.~Ferber,
Nucl.\ Phys.\ B {\bf 132}, 55 (1978).
\bibitem{isor1}
E.~A.~Ivanov and A.~S.~Sorin,
Sov.\ J.\ Nucl.\ Phys.\  {\bf 30}, 440 (1979) [Yad.\ Fiz.\  {\bf 30}, 853
(1979)].
\bibitem{Sorokin:1999jx}
D.~Sorokin,
Phys.\ Rept.\  {\bf 329}, 1 (2000) [arXiv:hep-th/9906142].
\bibitem{mar}
R.~Marnelius,
Phys.\ Rev.\ D {\bf 20}, 2091 (1979).
\bibitem{schwarz}
J.~H.~Schwarz,
Nucl.\ Phys.\ B {\bf 185}, 221 (1981).
\bibitem{sig}
W.~Siegel,
Int.\ J.\ Mod.\ Phys.\ A {\bf 3}, 2713 (1988).
\bibitem{abra}
E.~R.~Abraham, P.~K.~Townsend and P.~S.~Howe,
Class.\ Quant.\ Grav.\  {\bf 6}, 1541 (1989).
\bibitem{kuzenko}
S.~M.~Kuzenko and Z.~V.~Yarevskaya,
Mod.\ Phys.\ Lett.\ A {\bf 11}, 1653 (1996) [arXiv:hep-th/9512115].
\bibitem{pashnev}
A.~Pashnev,
Nucl.\ Phys.\ Proc.\ Suppl.\  {\bf 102}, 240 (2001) [arXiv:hep-th/0104249].
\bibitem{bars1}
I.~Bars, ``2T physics 2001'', hep-th/0106021 (and references therein).
\bibitem{bogo}
N.~N.~Bogolyubov and D.~V.~Shirkov, ``Introduction to the theory of Quantum
Fields'', John Wiley $\&$ Sons, NY, 1979.
\bibitem{gsw}
M.~B.~Green, J.~H.~Schwarz and E.~Witten, ``Superstring theory'', V. 1., CUP
1988.
\bibitem{rudichev}
I.~Rudychev and E.~Sezgin, Phys. Lett. B {\bf 415}, 363 (1997); Addendum-ibid.
B {\bf 424}, 411 (1998) [arXiv:hep-th/9704057];
Phys. Lett. B {\bf 424}, 60 (1998), arXiv:hep-th/9711128.
\bibitem{bl}
I.~Bandos and J.~Lukierski,
Mod.\ Phys.\ Lett.\ A {\bf 14}, 1257 (1999) [arXiv:hep-th/9811022].
\bibitem{isor} E.~A.~Ivanov and A.~S.~Sorin,
J.\ Phys.\ A {\bf 13}, 1159 (1980).
\bibitem{bls}
I.~Bandos, J.~Lukierski and D.~Sorokin,
Phys.\ Rev.\ D {\bf 61}, 045002 (2000) [arXiv:hep-th/9904109].
\\
I.~Bandos, J.~Lukierski, C.~Preitschopf and D.~Sorokin,
Phys.\ Rev.\ D {\bf 61}, 065009 (2000) [arXiv:hep-th/9907113].
\bibitem{fronsdal}
C.~Fronsdal,
Phys.\ Rev.\ D {\bf 12}, 3819 (1975).
\bibitem{abers}
E.~S.~Abers and B.~W.~Lee,
Phys.\ Rept.\  {\bf 9}, 1 (1973).
\bibitem{bastia}
F.~Bastianelli,
Nucl.\ Phys.\ B {\bf 376}, 113 (1992) [arXiv:hep-th/9112035].
\bibitem{parker}
L.~Parker,
Phys.\ Rev.\ D {\bf 19}, 438 (1979).
\bibitem{kuchar}
K.~Kuchar,
J.\ Math.\ Phys.\  {\bf 23}, 1647 (1982).
\bibitem{strominger}
J.~Michelson and A.~Strominger,
Commun.\ Math.\ Phys.\  {\bf 213}, 1 (2000) [arXiv:hep-th/9907191];
JHEP {\bf 9909}, 005 (1999) [arXiv:hep-th/9908044].
\\
R.~Britto-Pacumio, J.~Michelson, A.~Strominger and A.~Volovich, ``Lectures on
superconformal quantum mechanics and multi-black hole  moduli spaces'',
hep-th/9911066.
\\
G.~Papadopoulos,
Class.\ Quant.\ Grav.\  {\bf 17}, 3715 (2000) [arXiv:hep-th/0002007].
\bibitem{45}
J.~W.~van Holten and A.~Van Proeyen, J. Phys. A {\bf 15}, 3763 (1982).
\bibitem{46}
C.~Chryssomalakos, J.~A.~de Azc\'arraga, J.~M.~Izquierdo and J.~C.~Perez Bueno,
Nucl. Phys. B {\bf 567}, 293 (2000) [arXiv:hep-th/9904137].
\bibitem{zima}
S.~Fedoruk and V.~G.~Zima,
Nucl.\ Phys.\ Proc.\ Suppl.\  {\bf 102}, 233 (2001) [arXiv:hep-th/0104178].
\bibitem{BAIL}
I.~A.~Bandos, J.~A.~de Azc\'arraga, J.~M.~Izquierdo and J.~Lukierski,
Phys.\ Rev.\ Lett. {\bf 86}, 4451 (2001) [arXiv:hep-th/0101113].
\bibitem{vasiliev}
M.~A.~Vasiliev, ``Conformal higher spin symmetries of 4D massless
supermultiplets and  osp(L,2M) invariant equations in generalized
(super)space'', hep-th/0106149.
\\
M.~A.~Vasiliev, ``Relativity, causality, locality, quantization and duality in
the Sp(2M)  invariant generalized space-time'', hep-th/0111119.
\\
M.~A.~Vasiliev, ``Higher Spin Conserved Currents in $Sp(2M)$ Symmetric
Spacetime'', hep-th/0204167 (and references therein).
\end{thebibliography}
\end{document}